\documentclass[preprint]{aastex}

\usepackage{graphicx}
\usepackage{enumerate}
\usepackage{amsmath}

\def\mtsz{\scriptsize}
\def\orchestra{\textit{Orchestra}}

\newcommand{\lae}{\lower 2pt \hbox{$\, \buildrel {\scriptstyle <}\over {\scriptstyle\sim}\,$}}
\newcommand{\gae}{\lower 2pt \hbox{$\, \buildrel {\scriptstyle >}\over {\scriptstyle\sim}\,$}}

\def\eqnum[#1]{(\ref{#1})}
\def\mysubsubsection#1{\subsubsection{#1}}

\def\Msolar{\ifmmode {\rm M_{\odot}}\else $\rm M_{\odot}$\fi}
\def\Rsolar{\ifmmode {\rm R_{\odot}}\else $\rm R_{\odot}$\fi}
\def\Mearth{\ifmmode {\rm M_{\oplus}}\else $\rm M_{\oplus}$\fi}
\def\Rearth{\ifmmode {\rm R_{\oplus}}\else $\rm R_{\oplus}$\fi}

\def\Mstar{M_{\rm \ast}}
\def\Rstar{R_{\rm \ast}}
\def\Bstar{B_{\rm \ast}}
\def\Tstar{T_{\rm \ast}}

\def\magmostar{\vec{m}_{\rm \ast}}
\def\er{\hat{e}_r}
\def\ez{\hat{e}_z}

\def\Rroche{R_{\rm Roche}}

\def\Rm{{R}_m}

\def\rhop{\bar{\rho}}

\def\ropt{r_{\!\text{opt}}}

\def\Pohm{P_\text{Ohm}}
\def\Prad{P_\text{rad}}
\def\Pavg{\left<\Pohm\right>}
\def\rp{r_{\!\text{ast}}}
\def\mp{m}
\def\rinside{r}

\def\efffac{{\cal F}}

\def\rc{r_{c}}

\def\cpc{C_{p}}

\def\fmantle{f_{m}}

\def\Tcurie{T_{\rm Curie}}
\def\murel{\mu_{\rm rel}}
\def\muref{5000}   
\def\muvac{\mu_0}
\def\cpm{C_{p,m}}
\def\rhom{\rho_{m}}
\def\alpham{\alpha_{m}}

\def\Ohmic{Ohmic}

\begin{document}

\title{\textbf{\large \Ohmic\ heating of asteroids around magnetic stars}}

\author{Benjamin C. Bromley}
\affil{Department of Physics \& Astronomy, University of Utah, 
\\ 115 S 1400 E, Rm 201, Salt Lake City, UT 84112}
\email{bromley@physics.utah.edu}

\author{Scott J. Kenyon}
\affil{Smithsonian Astrophysical Observatory,
\\ 60 Garden St., Cambridge, MA 02138}
\email{skenyon@cfa.harvard.edu}

\begin{abstract}

We consider the impact of electromagnetic induction and
\Ohmic\ heating on a conducting planetary object that orbits a
magnetic star.  Power dissipated as heat saps orbital energy.  If this
heat is trapped by an insulating crust or mantle, interior
temperatures increase substantially.  We provide a quantitative
description of this behavior and discuss the astrophysical scenarios
in which it might occur.  Magnetic fields around some main-sequence
stars and white dwarfs are strong enough to cause the decay of
close-in orbits of asteroids and dwarf planets, drawing them through
the Roche limit on Myr time scales.  We confirm that Ohmic heating
around neutron stars is driven by the rotation of the stellar magnetic
dipole, not orbital dynamics.  In any case, heating can raise interior
temperatures of asteroids or dwarf planets on close-in orbits to well
above liquidus.  Hot material escaping to the surface may lead to
volcanic ejections that can obscure the host star (as in the light
curve of KIC 8462852) and pollute its atmosphere (as observed with
metal-rich white dwarfs).  We speculate that mixing of a volatile-rich
mantle or crust with material from an induction-heated core may lead
to an explosion that could destroy the asteroid prior to tidal
break-up.

\end{abstract}



\section{Introduction}

A magnetic star is a powerhouse. An asteroid passing close by
experiences magnetic flux variations that drive an internal electric
current.  The interplay between this current, the magnetic field, and
the asteroid's conductivity establishes how deep the magnetic flux 
penetrates into the asteroid. Resistive losses are inevitable.  In the 
reference frame of the star, these losses come at the expense of the 
asteroid's kinetic energy.  As the asteroid is heated, its orbit decays.

To explore this scenario, we rely on a quantitative description of the
interaction between an idealized conducting sphere and a time varying
magnetic field. Solutions to this problem have appeared in the
literature for a wide range of applications.  \citet{bidinosti2007}
provide a nice bibliographic overview, pointing out the early work of
Mie, Debye (in scattering problems), Wait (geophysics), and Hoult and
Lauterbur (NMR). More recently \citet{ray2018} add in the effect of
magnetic field gradients to generate a net force on the conductor as
a method for sorting recyclable metals.

The ingredients for Ohmic heating are available in astrophysical
settings.  Iron or iron-nickel asteroids like (16)~Psyche and planets
with metallic cores like the Earth have regions of high conductivity.
Hydrous or rocky planets may also be electrically conductive and able
to sustain internal currents from electromagnetic induction.  Examples
of this phenomenon, sometimes called induction heating or Joule
heating, include a conducting asteroid plowing through the
magnetized plasma surrounding the young Sun or a T~Tauri star
\citep{sonett1970, mckinnon1989b, shimazu1995, menzel2013}, unipolar
induction through the moons of Jupiter and Saturn
\citep{piddington1968, goldreich1969, hand2011}, and the heating of
asteroids by the time-varying magnetic field of a pulsar
\citep{cordes2008, kotera2016}. Other applications include heating of
conducting zones within hot Jupiters \citep{batygin2011, wu2013,
  ginzburg2016b} and accretion flows
\citep[e.g.][]{bisnovatyi-kogan1997, okuzumi2013, khaibrakhmanov2018}.

We focus here on a distinct mechanism for \Ohmic\ heating driven by
electromagnetic induction. When an asteroid orbits a magnetic star, it
typically experiences a changing magnetic flux even if the stellar
magnetic field is static in the star's rest frame.  We expect strong
flux variations for highly eccentric orbits when the periastron is
near the Roche limit, as when a comet or asteroid is scattered toward
the stellar host by a more distant planet \citep[e.g.,][]{rasio1996,
  chatterjee2008}.  The resulting electromotive force drives
electrical currents within a conducting asteroid, generating
\Ohmic\ heat. The net magnetic force on the eddy currents produces a
drag, causing a loss of speed and orbital energy. The amount of energy
drawn per orbit, as we quantify below, is small compared to the energy
in the star's magnetic field. Thus, any back-reaction of the
asteroid's slow orbital evolution onto the star and its magnetic
dynamo is inconsequential; all of the energy dissipated by induced
currents comes from the loss of orbital energy.

\Ohmic\ dissipation in a conducting asteroid contributes to its
overall thermal evolution. \Ohmic\ heating takes place in specific
regions of the asteroid, for example only in a conducting core, and
then only near the core's surface if the skin effect limits the
magnetic penetration into the conducting medium. \Ohmic\ heat thus
serves as a localized energy source, contributing to the heat flow
throughout the asteroid. Other factors include stellar illumination
and radiative cooling at the surface, along with the asteroid's
physical and material characteristics. Thermal time scales are rapid
compared to orbital evolution; small asteroids may reach a steady
state quickly, with energy input from starlight and \Ohmic\ heating
balanced by radiative cooling. A larger asteroid with a conducting
core that is thermally insulated by a deep mantle may undergo runaway
core heating. The diffusion equation quantifies this process, helping
to distinguish among hypothetical possibilities.

These outcomes require close-in orbits around stars with strong
magnetic fields.  Magnetic pre-main-sequence stars and some late-type
dwarfs have fields as strong as $\Bstar \sim 10^3$~G on the stellar
surface \citep[e.g.,][]{johns-krull2009}.  Some Ap stars, such as
HD~215441 \citep{babcock1960} and HD~75049 \citep{elkin2010}, have
much stronger fields ($\Bstar \gtrsim 30$~kG). A few magnetic white
dwarfs have surface field strengths approaching $10^9$~G
\citep[e.g.][]{angel1981, schmidt1995, wickramasinghe2000,
  ferrario2015}. Neutron stars can achieve field strengths as high as
$10^{15}$~G (\citealt{kaspi2017} provide a review). Close-in orbits
around these stars are also plausible, as they have been observed
directly in exoplanet searches (\citealt{sanchis2014}; KOI 1843.03 has
the shortest known orbital period, 4.245~hr around a red dwarf;
\citealt{ofir2013}; \citealt{rappaport2013}). Close-in planetary
bodies are also inferred from the presence of metals in white dwarf
atmospheres \citep[e.g.,][]{jura2006, farihi2010} as well as
transiting dust clouds \citep[e.g., WD 1145+017;][]{alonso2016}.

Here we consider if electromagnetic induction within orbiting bodies
can sap orbital energy at a rate large enough and generate enough heat
to have observable consequences in real astrophysical systems.  First,
we describe electromagnetic induction in a spherical conductor
(\S\ref{sec:induction}), and the impact of induced currents on its
orbit (\S\ref{sec:orbit}) and its thermal evolution
(\S\ref{sec:heat}). Then we discuss how this phenomenon might play out
in several astrophysical scenarios (\S\ref{sec:discuss}). We
summarize our results (\S\ref{sec:conclude}).

\section{Electromagnetic induction}\label{sec:induction}

An electrically conducting asteroid, or one with a conducting core,
experiences an oscillating magnetic field as it orbits a magnetic
star. To calculate the \Ohmic\ heating from eddy currents that draw
energy from orbital motion, we first consider the case of an 
idealized conductor sitting at rest in a time varying magnetic
field. This situation gives the response of the asteroid to one
Fourier mode of the time-varying field, as seen in the asteroid's 
rest frame.  We then apply the results to the situation where time
variations correspond to full orbital motion through the stellar
field.

\subsection{\Ohmic\ heating in a time-varying uniform magnetic field}

In an idealized model, an asteroid is a sphere of radius $\rp$ and
mass $\mp$, with constant, real-valued electrical conductivity
$\sigma$ and magnetic permeability $\mu$.  We assume that $\rp$ is
small, so that the stellar magnetic field $\vec{B}$ in its vicinity
is, to a good approximation, spatially uniform. Then, $\vec{B} = B_0
\exp(-i\omega t) \ez$, where the field is aligned with the $z$-axis
in the asteroid's reference frame, and $\omega$ is an angular frequency
characteristic of orbital motion. This frequency is low enough
that we may safely ignore electromagnetic scattering
\citep[$\omega\epsilon \ll \sigma$, where $\epsilon$ is the
  permittivity; see][]{kotera2016}. The induced electric field
$\vec{E}$ and current density $\vec{J}$ are related through Ohm's Law
($\vec{J} = \sigma \vec{E}$); thus, we can use Maxwell's equations to
solve for $\vec{B}$ and $\vec{J}$ directly, deriving the
\Ohmic\ heating from $|\vec{J}|^2/\sigma$.

Motivated by research in magnetic resonance imaging,
\citet{bidinosti2007} obtain an elegant solution for Ohmic power loss
in the problem of a homogeneous conducting magnetic sphere in a time
varying magnetic field, extending earlier work that considered
non-magnetic media. In their application, the field variations are
rapid (radio frequencies) and the spheres are small (centimeters), but
their solution is general, applying equally well to orbital time
scales and kilometer-size objects.\footnote{We follow
  \citet{bidinosti2007} in adopting the SI system for Maxwell's
  Equations and derived quantities. We use mixed units when giving
  characteristic values of these and related quantities, according to
  typical uses in the literature.}  By matching boundary conditions
for the magnetic field at large distances and at the sphere's surface,
they show that the current density within the asteroid lies in the
azimuthal direction with magnitude
\begin{equation}\label{eq:J}
  J(\rinside,\theta) = \frac{3 \pi B_0 |k|^2 \rinside 
    \left[ j_0(k\rinside) + j_2(k\rinside)\right]%
\sin(\theta)}{2 \left[ (\mu+2\muvac) j_0(k\rp) + (\mu-\muvac) 
j_2(k\rp)\right]},
\end{equation}
where $\rinside$ and $\theta$ are spherical polar coordinates with an origin
at the asteroid's center,
$k = \sqrt{i\mu\sigma\omega}$, $\muvac$ is the vacuum
permeability,
and
\begin{eqnarray}
j_0(z) &  = & \sin(z)/z \\
j_2(z) &  = &  (3/z^3-1/z)\sin(z) - 3\cos(z)/z^2 
\end{eqnarray}
are spherical Bessel functions.\footnote{When evaluating the spherical
  Bessel functions with complex arguments, we use the forms with sines
  and cosines, for which modern programming languages (\texttt{python}
  and \texttt{C++}) support complex arithmetic.}  In this Fourier
analysis, the (complex) current density oscillates as $\exp(-i\omega
t)$.

A quantity that characterizes the current density is the
skin depth $\delta = \sqrt{2/\mu\sigma\omega}$, which gives an
indication where in the sphere the current is strong. In the limit of
low frequency, for example, $\delta$ can be large, exceeding the
conductor's size. Then the current is spread over much of the
asteroid's interior.  At high frequencies, $\delta$ is small, and
current is confined to the surface of the sphere.

The average power lost to \Ohmic\ heating in the asteroid is
\begin{equation}\label{eq:POhm}
 \Pohm = 
\frac{1}{2} \int_\text{sphere}\!\! \frac{|J(\rinside,\theta)|^2}{\sigma} dV 
 \  = \ \frac{2 \pi\rp^2 B_0}{\muvac}\,\text{Im}\left\{\frac{J(\rp,\pi/2)}{i\sigma}\right\},
\end{equation}
where the rightmost expression, from \citet[Eq.~20
  therein]{bidinosti2007}, is the product of the magnetic flux through
the sphere's midplane and the electric field strength at its equator.
\citet{bidinosti2007} identified it for the case of non-magnetic media
($\mu = \muvac$), and we confirm that the expression holds for
magnetic material ($\mu > \muvac$). With this result we can write the
power loss as the product of the average magnetic energy density, the
rate at which the magnetic field changes, the volume of the sphere,
and an ``efficiency factor'' $\efffac$ that depends only on dimensionless
quantities:
\begin{eqnarray}\label{eq:P}
\Pohm  & = &\frac{B_0^2}{2\muvac} \ \omega \ 
\frac{4\pi\rp^3}{3} \  
\efffac(\Rm,\murel)
\\
\efffac(\Rm,\murel) & \equiv &
\text{Im}
\left\{ 
\frac{ 9
\murel \left[j_0(z) + 
j_2(z)\right]}%
{ 2 
[(\murel + 2) j_0(z) + (\murel-1) j_2(z)]}
\right\},
\end{eqnarray}
where $\murel\equiv \mu/\muvac$ and $\Rm = \mu\sigma\omega\rp^2$ are
the relative permeability and magnetic Reynolds number, respectively,
and $z \equiv \sqrt{i\Rm}$.  The Reynolds number is related to the
skin depth through $\delta = \rp\sqrt{2/\Rm}$. In general, $0 \leq
\efffac < 1$.

\begin{figure}[htb]
\centerline{\includegraphics[width=5.0in]{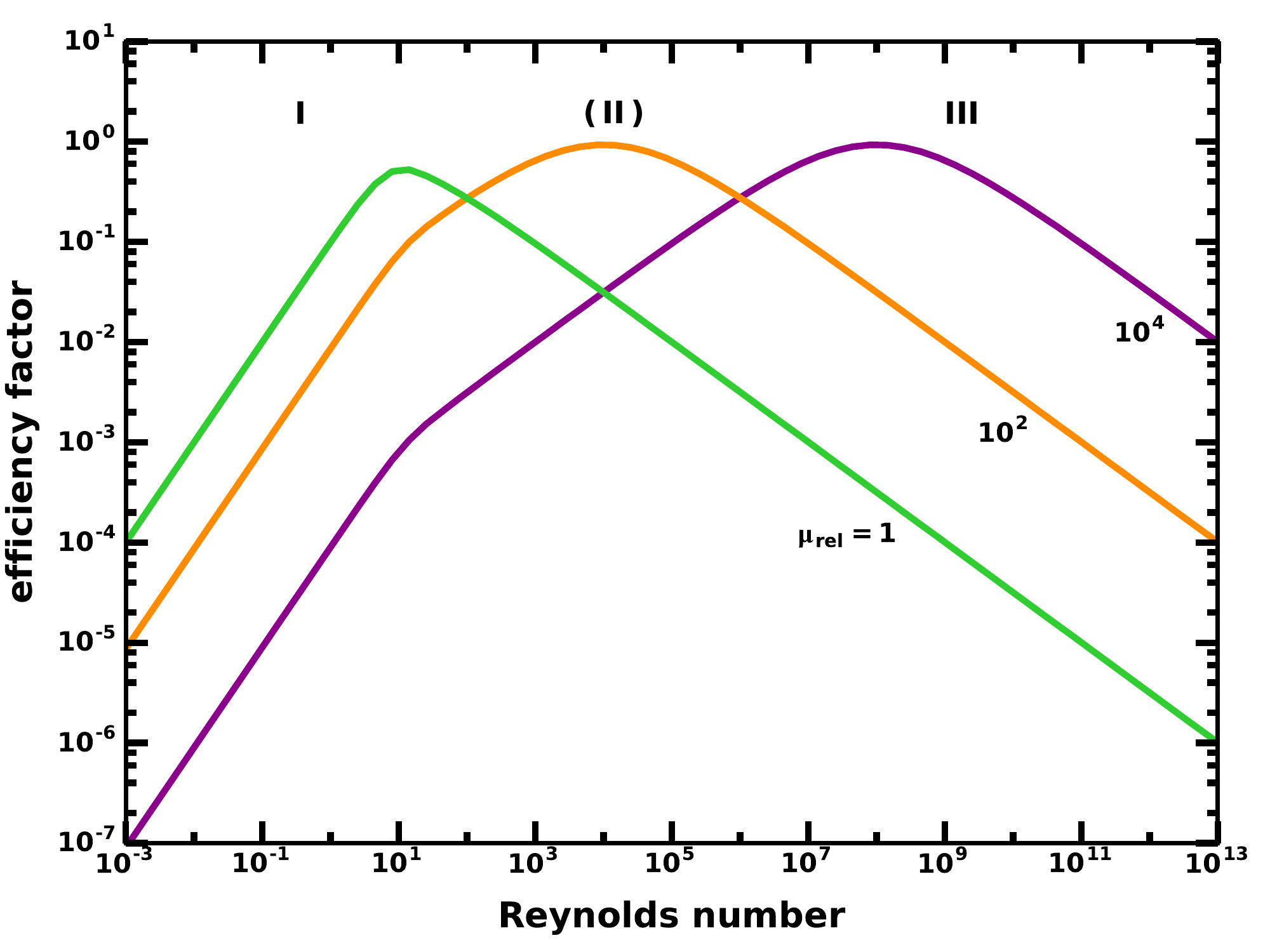}}
\caption{\label{fig:ploss} The efficiency factor as a function of
  magnetic Reynolds number for a spherical conductor.  At low $\Rm$,
  labeled as regime I, the applied (stellar) magnetic field dominates
  throughout the medium and the current density builds up in the bulk
  of the conductor. The efficiency factor varies as $\Rm^2$.  In
  region II, applicable only to magnetic media with relative
  permeability $\murel>1$, the efficiency factor scales as $\Rm^{1/2}$
  as a result of the contributions of the magnetization in response to
  the changing magnetic flux. Regime III corresponds to $\Rm >
  \murel^2$.  There the efficiency factor declines as $\Rm^{-1/2}$. In
  this regime, the surface current eliminates the applied
  field inside the conductor.  }
\end{figure}

The efficiency factor $\efffac(\Rm, \murel)$ encodes the detailed
physics of electromagnetic induction and Ohmic heating. Its behavior
falls into three distinct regimes, depending on the values of the
Reynolds number and the relative permeability:
\begin{enumerate}[I.]
\item When the Reynolds number is low, $\Rm \lesssim 3$, the
  efficiency factor scales as $\efffac \sim \Rm^2$. The local stellar
  magnetic field permeates through the conducting asteroid, dominating
  over the induced field. The skin depth is comparable to the radius
  of the sphere or larger, indicating that an eddy current forms
  within the bulk material.
\item At intermediate Reynolds number, $3 \lesssim \Rm \lesssim
  \murel^2$, the efficiency factor scales as $\efffac \sim \Rm^{1/2}$. 
  The current density is concentrated near the asteroid's surface but 
  magnetization from the bulk material also responds to the applied 
  field. This regime applies only to magnetic material.
\item At high Reynolds number, $\Rm \gtrsim \murel^2$, the efficiency
  factor falls off as $\efffac \sim \Rm^{-1/2}$. The current density is
  nearly an ideal surface current, $K = J\delta$, which cancels the
  varying magnetic field in the interior of the asteroid independently
  of any bulk magnetization. 
\end{enumerate}
If the conducting medium is non-magnetic ($\murel \sim 1$), then the
efficiency transitions from regime I directly to regime III.  The
efficiency factor reaches a peak of approximately 53\%\ at the
transition point, near $\Rm = 11.6$. In high-permeability material
($\murel\gg 1)$, the efficiency factor has a maximum that is just
above 93\% at $\Rm = \murel^2$.  Figure~\ref{fig:ploss} provides an
illustration, showing efficiency curves ($\efffac$ versus $\Rm$) for
several values of $\murel$.

Expressions of the power loss in the three regimes are
\begin{align}
\label{eq:PI}
P_{I}  & \approx \ 
\frac{3\pi}{5}\, B^2 \murel^2 (\murel+2)^{-2}
\sigma \omega^2
\rp^5  
\ \ \ \ 
&& (\Rm \lesssim 3)
\\
\label{eq:PII}
P_{II}  & \approx \ \frac{2}{3}\, B^2 
\, (\muvac\murel)^{-1/2}\,
\sigma^{1/2}
\omega^{3/2} \rp^{4} 
&& (3 \lesssim \Rm \lesssim \murel^2) 
\\
\label{eq:PIII}
P_{III} & \approx \
\vphantom{\frac{5}{3}}
3\pi\,B^2 
\,(\murel/\muvac^3)^{1/2}\,
\sigma^{-1/2}
\omega^{1/2}
\rp^2 
&& (\Rm \gtrsim \murel^2).
\end{align}

With ferromagnetic material, a complicating factor is that
the relative permeability drops to unity when the temperature
increases beyond the Curie temperature $\Tcurie$, around 10$^3$~K for pure
iron. We discuss this effect below. We do not consider other phenomena
associated with ferromagnets, such as hysteresis.

If we fix the bulk material properties of an asteroid, along with the 
frequency $\omega$, we can identify an optimal radius that maximizes 
the efficiency $\efffac$: 
\begin{align}
\label{eq:ropt}
\ropt \sim 
\left\{
\def\arraystretch{1.2}
\begin{array}{ll}
3.4/\sqrt{\muvac\sigma\omega} \ \ \ \ & (\murel \sim 1) \\
\sqrt{\murel/\muvac\sigma\omega} \ \ \ \ &  (\murel \gg 1),
\end{array}
\right.
\end{align}
The upper equation is appropriate to solids made of ice, water or
rock, while the lower equation is for solid ferromagnetic asteroids or
their cores. In all cases, the frequency $\omega$ corresponds to an
oscillatory mode of the magnetic field. Multiple modes may contribute
to the heating process; which ones are determined by the time
variation in the magnetic field as the asteroid orbits its stellar
host. 

As an example, a metallic solid orbiting in the dipole field of the Sun
at 0.1~AU, the optimal radius for \Ohmic\ heating is roughly a
kilometer. For a rocky asteroid, $\ropt$ can exceed a 100~km. 
The precise value of $\ropt$ depends on material properties, the nature 
of the stellar magnetic field, and how an orbiting body moves through 
the field. These topics are next.

\subsection{Orbital motion in the dipole magnetic field}%
\label{subsec:induction}

When the stellar field is a dipole, the magnetic field at a position
$\vec{d}$ relative to the star is
\begin{equation}
\vec{B} = \Bstar \frac{\Rstar^3}{d^3} \left[ 3 (\er\cdot\ez) \er - \ez
  \right],
\end{equation}
where $\Bstar$ specifies the field strength at the surface of the star
at radius $\Rstar$, and the unit vector $\ez$ is aligned with the
star's magnetic dipole moment $\magmostar$. We assume that the
magnetic field in the star's rest frame is static, with fixed
orientation. We further designate that a reference orbital plane lies
perpendicular to $\magmostar$; orbits in this plane have inclination
$i = 0^\circ$.  Additionally we assume that the effect of the magnetic
field on an orbiting solid is small compared to gravity, so that its
osculating orbit is Keplerian with mean angular velocity $\Omega_a =
\sqrt{G\Mstar/a^3}$, where $\Mstar$ is the stellar mass and $a$ is the
semimajor axis.

The idealization of a static dipole field carries the assumption that
the asteroid's orbital motion is fast compared with the rotation of
the star or that the dipole moment is closely aligned with the star's
angular momentum (or both). While this approximation is not valid for
pulsars, it is reasonable for main-sequence stars and most white
dwarfs.

In the asteroid's reference frame, a static magnetic field is
perceived as time varying, provided that the asteroid has either some
eccentricity $e>0$ or inclination $i>0^\circ$. Two limiting cases are
(i) a circular ``polar'' orbit of radius $a$ in a plane containing the
stellar magnetic dipole moment ($i=90^\circ$) and (ii) a highly
eccentric ``equatorial'' orbit in the fiducial orbital plane
($i=0^\circ$). In the first case, the magnetic field strength
variations are predominantly sinusoidal, with an angular frequency
that is twice the Keplerian value $\Omega_a$, since north and south
poles are equivalent in terms of the strength of the changing magnetic
field. In this preliminary work, we focus only on the field strength,
ignoring changes in the alignment of the magnetic field.  The
approximate orbit-averaged power loss is then given by
Equation~(\ref{eq:P}) with $\omega = 2\Omega_a$.

In the case of an eccentric orbit with semimajor axis $a$ and
periastron distance $q$, we estimate the \Ohmic\ power loss by first
approximating the inner part of an asteroid's orbit using a parabolic
trajectory with the same periastron distance.  We then decompose the
magnetic field sampled along this route into its frequency components
with an FFT of $B(d(t))$ over a time interval $T$, set to be about 10
times the circular orbital period at the periastron distance.  We find
that the spectrum of $B(\omega)$ is well-approximated as an
exponential with decay constant $v_q/q$, where $v_q$ is the asteroid's
speed at periastron. Accordingly, we estimate power loss during
periastron passage using Equation~(\ref{eq:P}) at characteristic
frequency $\omega = \sqrt{2}\Omega_q \approx v_q/q$, where $\Omega_q =
\sqrt{G\Mstar/q^3}$ is the angular frequency of a circular orbit at
distance $q$. Multiplying this power loss by the duration of passage,
$2q/v_q$, yields the energy loss per orbit. Comparison with
calculations using the full spectrum suggests that this approximation
is useful when the eccentricity of the orbit is $e \gtrsim 0.4$.

To summarize the results just described, the \textit{orbit-averaged}
\Ohmic\ power loss experienced by an asteroid is
\begin{gather}\label{eq:Pavg}
\Pavg \approx 
\frac{2\rp^3}{3}\,\frac{\Bstar^2}{\muvac} \, 
\Omega_a
\, \efffac(\Rm,\murel; \omega) \, \times \,
\left\{
\def\arraystretch{1.1}
\begin{array}{ll}
2\pi \Rstar^6/a^{6}\ \ \ \ 
& (\text{circular polar; $\omega = 2\Omega_a$})  
\\
\Rstar^6/q^{6}
& (\text{eccentric, $0.4 \lesssim e  < 1$;  $\omega = \sqrt{2}\Omega_q$}),
\end{array}
\right.
\end{gather}
where the Reynolds number is evaluated with the angular frequency
$\omega = 2\Omega_a$ and $\sqrt{2}\Omega_q$ for circular polar orbits
($i=90^\circ$) and eccentric equatorial orbits ($i = 0^\circ$),
respectively. In general, objects with a size near $\ropt$
(Eq.~(\ref{eq:ropt})) experience the greatest amount of heating per
unit mass.  We focus first on these objects when considering whether
``orbit-induced'' \Ohmic\ heating might play a role in astrophysical
scenarios.

\section{Orbital evolution}\label{sec:orbit}

An asteroid moving through the magnetic field of its stellar host
feels a Lorentz force as a result of interactions between the induced
currents within it and the stellar field.  While the detailed dynamics
can be complicated, depending on the nature of the field gradients and
the orbit geometry, our interest is on how the overall orbit changes
with time.  Our focus is specifically on orbital energy, as it is most
connected to \Ohmic\ heating.

For the effects described here to be important, an asteroid must
transit rapidly through a region of strong magnetic field, close to
its host. If the asteroid is to survive the stellar tidal forces
during the encounter, it must not venture too far inside the Roche
limit,
\begin{equation}
\label{eq:rroche}
  \Rroche  = k \left(\frac{\Mstar}{\rhop}\right)^{1/3}  
 \approx 0.86 k
\left[\frac{\Mstar}{\text{\mtsz 0.5~$\Msolar$}}\right]^{1/3}
\left[\frac{\rhop}{\text{\mtsz 5 g/cm$^3$}}\right]^{-1/3}
\ \text{\Rsolar},
\end{equation}
where $\rhop$ is the average mass density of the conducting asteroid
and the parameter $k$ is a constant of order unity.  A value of
$k=0.8$ corresponds to the case where the asteroid is a strengthless,
non-rotating solid body. Rotating objects have larger $k$, as do fluid
bodies; objects with significant tensile strength have smaller $k$
\citep[e.g.,][see also \citealt{davidsson1999, jura2003,
    holsapple2008}]{veras2017}.  

For the value of $k=1$ adopted here, the Roche radius lies outside of
the stellar surface for stars that are more dense than the Sun,
including red dwarfs, white dwarfs and neutron stars. For less dense
stars with $\Rroche < \Rstar$, which include main-sequence stars that
are more massive than the Sun, we consider star-grazing orbits
instead.

With the Roche limit (Eq.~(\ref{eq:rroche})) as a guide, we estimate
the optimal physical radius of an asteroid for efficient \Ohmic\ heating
(Eq.~(\ref{eq:ropt}). Around a late-type star or white dwarf,
\begin{align}
\label{eq:roptorbit}
\ropt \  & \sim \ 
\vphantom{\left[\frac{\pi}{\pi}\right]^{\pi}}
\left(\frac{\murel}{\muvac\sigma}\right)^{1/2} 
\left(\frac{q^3}{G\Mstar}\right)^{1/4}
&&
\ 
\\ \label{eq:roptmetal}
\ & \approx 
\hphantom{006}3 
\left[\frac{\murel}{\text{\mtsz \muref}}\right]^{\!1/2}\!
\left[\frac{\sigma}{\text{\mtsz $10^6$~S/m}}\right]^{\!-1/2}\!
\left[\frac{\Mstar}{\text{\mtsz 0.5 $\Msolar$}}\right]^{\!-1/4}\!
\left[\frac{q}{\text{\mtsz 1 \Rsolar}}\right]^{\! 3/4} 
\ \text{km} \ \ \ 
&& (\text{iron/nickel}),
\\ \label{eq:roptwater}
\ & \approx 
\hphantom{00}64 
\left[\frac{\sigma}{\text{\mtsz 5~S/m}}\right]^{\!-1/2}\!
\left[\frac{\Mstar}{\text{\mtsz 0.5 $\Msolar$}}\right]^{\!-1/4}\!
\left[\frac{q}{\text{\mtsz 1 \Rsolar}}\right]^{\! 3/4} 
\ \text{km} \ \ \ 
&& (\text{water}),
\\ \label{eq:roptrock}
\ & \approx 
1400 
\left[\frac{\sigma}{\text{\mtsz $0.01$~S/m}}\right]^{\!-1/2}\!
\left[\frac{\Mstar}{\text{\mtsz 0.5 $\Msolar$}}\right]^{\!-1/4}\!
\left[\frac{q}{\text{\mtsz 1 \Rsolar}}\right]^{\! 3/4} 
\ \text{km} \ \ \ 
&& (\text{rock}).
\end{align}
The numerical values of $\murel$ and $\sigma$ are order-of-magnitude
estimates for solid iron in a ferromagnetic asteroid
\citep[e.g.,][]{dekoker2012} and liquid water in a ``water world''
\citep{thomas1934}.  The conductivity of rocky, planetary material can
vary enormously, from $\sim 10^{-5}$~S/m to $O(1)$~S/m
\citep[e.g.,][]{schwarz1990}. Here, we adopt an intermediate value of
$0.01$~S/m \citep[cf.][]{xu1998}.

The numerical reference values in
Equations~(\ref{eq:roptmetal})--(\ref{eq:roptrock}) also reflect the
angular frequency of magnetic flux variation, which arises from
orbital motion.  The reference values are based on a close-in,
circular orbit about a red dwarf host. If the orbit were eccentric,
but with the same periastron distance, the optimal radius would
increase. Still the main frequency component of the magnetic flux
variations would be similar, reflecting the fact that $\omega \sim
v/q$ where the field is strongest.  Thus, for a given periastron
distance $q$ we use the same $\ropt$ value when considering circular
and eccentric orbits.

Generally, reducing the orbital distance reduces $\ropt$.  The optimal
radius also shrinks as the stellar mass increases, so long as the
asteroid's orbit is close to star-grazing.

Next, we consider specific orbital configurations for which Ohmic
heating can play a role. Circular polar orbits perpendicular to the
magnetic dipole moment lead to inspiral, while eccentric orbits that
cut in and out of the field in the plane perpendicular to $\magmostar$
experience eccentricity damping at approximately fixed periastron
distance.  In most astrophysical situations, the relative orientation
of asteroid orbits falls somewhere between these two idealizations.

\subsection{Circular inspiral}\label{subsec:inspiral}

As \Ohmic\ heating draws orbital energy, the orbit of an asteroid at
semimajor axis $a$ decays at a rate of
\begin{equation}\label{eq:dadt}
\frac{da}{dt} = -\frac{2a^2\Pavg}{G\Mstar\mp},
\end{equation}
where the orbit-averaged \Ohmic\ heating rate $\Pavg$ is given in
Equation~(\ref{eq:Pavg}). To explore the possibilities of orbital
decay from \Ohmic\ heating, we focus on asteroids with radii
comparable to $\ropt$ (Eq.~(\ref{eq:roptorbit})), and on a circular
polar orbit near the Roche limit.  If the efficiency factor does not
change substantially as the asteroid spirals inward, $da/dt \sim
-a^{-11/2}$.  The formal solution to this ODE yields a time scale for
the inspiral, 
\begin{eqnarray}\label{eq:tinspiral}
\tau_\text{inspiral}\  & \ \approx \ &  \frac{1}{13}\, 
\rhop \, \left({G\Mstar}\right)^{1/2}\, 
\frac{\muvac\, a^{13/2}}{\Rstar^6\, \Bstar^2} 
\efffac^{-1}
\\ \label{eq:tinspiralexample}
\ & \ \approx \ & 0.84
\left[\frac{\rhop}{\text{\mtsz 5~g/cm$^3$}}\right]\!
\left[\frac{\Mstar}{\text{\mtsz 0.5~$\Msolar$}}\right]^{\!1/2}\!
\left[\frac{\Rstar}{\text{\mtsz 0.5~$\Rsolar$}}\right]^{\!-6}\!
\left[\frac{\Bstar}{\text{\mtsz 5~kG}}\right]^{\!-2}\!
\left[\frac{a}{\text{\mtsz 1 \Rsolar}}\right]^{\!13/2} 
\efffac^{-1}
\ \text{Myr},
\end{eqnarray}
where the power-law indices show how parameters scale for an asteroid
with radius near $\ropt$ and the numerical reference values are
plausible for a magnetic red dwarf.  This time scale suggests that
orbital decay by \Ohmic\ heating is possible in an astrophysical
setting.

This analysis carries the assumption that an asteroid remains close to
optimal size as its orbit evolves. In actuality, the efficiency factor
for \Ohmic\ heating changes as the body moves inward. Nonetheless, the
asteroid's inspiral rate will continue to get faster as inspiral
proceeds. For example, when $\efffac$ shifts away from its peak into
regime III (e.g., Fig.~\ref{fig:ploss}), the inspiral rate still increases,
with the orbital distance decreasing as $da/dt \sim -a^{-4}$.

The time scale for orbital decay in Equation~(\ref{eq:tinspiral}),
with numerical values tuned for a red dwarf, is similar for white
dwarfs and early-type magnetic stars. In all cases, the asteroid needs
to be within a Solar radius or so of its stellar host to experience
``magnetic migration''. Bodies larger than roughly a kilometer, which
are bound by gravity, have a Roche limit that is comparable to this
distance. They cannot inspiral much before being tidally disrupted.
Small bodies with high tensile strength ($\rp \ll 1$~km) can orbit
closer to the host star than larger ones \citep[e.g.,][]{kb2017b}. 
However, so close to the star there may be competition between
\Ohmic\ heating and other effects like radiative heating. 

\subsection{Eccentricity damping}

An asteroid on an eccentric orbit also loses orbital energy to Ohmic
heating. Most of this dissipation takes place near periastron, where
kinetic energy is lost with no significant deflection from a Keplerian
path. Thus, with each close passage, the asteroid slows incrementally,
keeping its periastron distance fixed while shaving off apoastron
distance at the expense of its semimajor axis. In this way the
asteroid's eccentricity damps as its semimajor axis falls.  In the
final stages of circularization of an asteroid in the equatorial
plane, the decay of the semimajor axis slows because the magnetic
field \textit{variations} encountered by the asteroid become less
significant, despite that the field itself is comparatively strong
there.  Then the asteroid damps onto a circular orbit just beyond the
initial periastron. If the asteroid has a non-zero inclination, it
will always encounter magnetic field variations along its orbit. It
thus settles onto a circular orbit and inspirals toward the host star.

To demonstrate this effect, we use the $n$-body routine in our
\orchestra\ code \citep[e.g.,][]{bk2011a}, modified to include the
loss of kinetic energy from \Ohmic\ heating. We choose an asteroid
that has a radius below optimal size, $\rp < \ropt$ ($\Rm \lesssim 1$;
regime I), so that the magnetic field is locally unperturbed by the
asteroid's eddy currents. The instantaneous force on the asteroid then
can be estimated directly from Maxwell's equations
\citep{vanbladel1988, giffin2010}, yielding
\begin{equation}\label{eq:F}
    \vec{F} = -\frac{2\pi\sigma\rp^5}{15}\vec{v} 
    \left|(\hat{e}_v\cdot\vec{\nabla})\vec{B}\right|^2 
    \ \ \ \ \ \ (\Rm \lesssim 1)
\end{equation}
where $\vec{v}$ is the asteroid's velocity, $\hat{e}_v \equiv
\vec{v}/v$, and $\vec{B}$ is the local value of the stellar magnetic
field.  This force acts as a drag, opposing the asteroid's velocity.
The strong dependence on radial distance ($d^{-17/2}$), as well as
orientation (the gradient in the magnetic field strength vanishes at
periastron when $i=0^\circ$), indicates that the force on the asteroid
varies widely along its orbit. If the orbital eccentricity is modestly
high, the asteroid gets the strongest kicks just before and after it
approaches periastron.

\begin{figure}
\centerline{\includegraphics[width=5.0in]{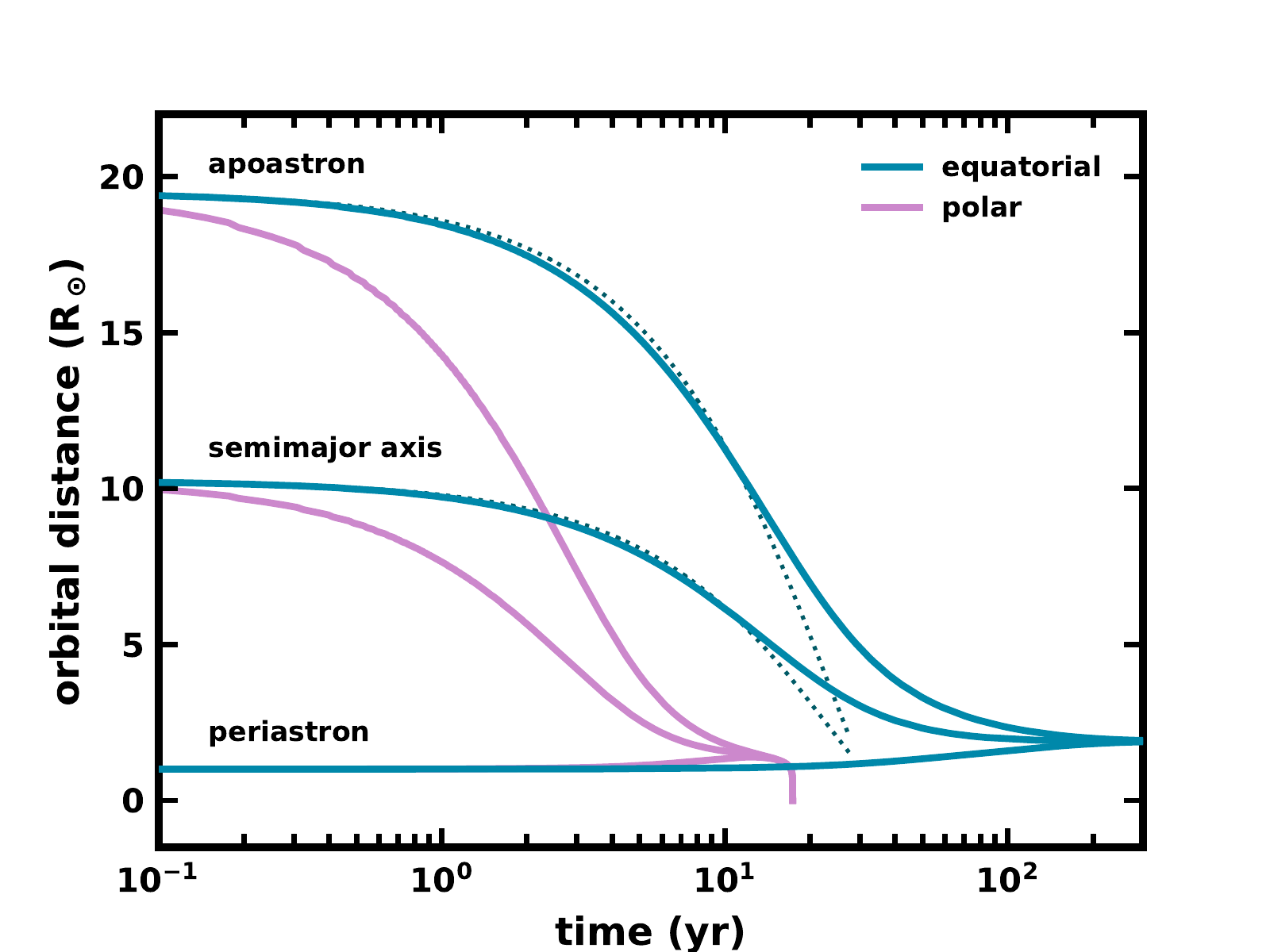}}
\caption{\label{fig:wdb} Simulation of a large non-magnetic rocky
  asteroid ($\rp = 250$~km, $\rho = 3$~g/cm$^3$, $\sigma = 0.01$~S/m),
  around a red dwarf ($\Rstar = 0.5$~\Rsolar, $\Mstar =
  0.5$~\Msolar) with an unrealistically high magnetic field ($\Bstar =
  5\times 10^7$~G) to reduce the computational run time needed to
  observe orbital evolution. The blue-green curves are for an orbit in
  the equatorial plane (perpendicular to the stellar magnetic dipole),
  the solid curve is from a simulation while the dotted curve is the
  theoretical expectation.  The eccentricity damps at nearly fixed
  periastron until the asteroid's orbit gets close to
  circular. Because the magnetic field is constant at fixed orbital
  distance, this asteroid will settle on circular orbit.  The magenta
  curves are for an asteroid on a polar orbit; the stronger field
  variations experienced by the asteroid on that orbit yield a faster
  damping time. Furthermore, those field variations remain even after
  circularization, hence the rapid plummet in the orbital distance
  once the eccentricity reaches zero.}
\end{figure}

Figure \ref{fig:wdb} illustrates orbital evolution with the force law
in Equation~(\ref{eq:F}) using \orchestra. The figure shows
periastron, semimajor axis, and apoastron of a 250~km rocky asteroid
on a close-in eccentric orbit around a red dwarf host. Here, we
artificially increase the stellar magnetic field by four orders of
magnitude above observed values for red dwarfs so that we limit the
number of orbits we need to integrate to measure the orbital
evolution. In the figure, we see that eccentricity damping is a
general feature, independent of the asteroid's orbital inclination.
However, if the asteroid is in the equatorial plane, its orbit stops
evolving after it circularizes --- the magnetic field is constant in
its frame, and there is no \Ohmic\ heating. If the orbit is inclined,
then the asteroid always experiences some field variations; circular
inspiral is inevitable, as in \S\ref{subsec:inspiral}.

Here we focus on the orbital evolution of a highly eccentric asteroid,
before it circularizes. The average \Ohmic\ power loss of an asteroid
on an eccentric orbit at fixed periastron distance scales with
semimajor axis as $a^{-3/2}$. This result stems from
Equation~(\ref{eq:Pavg}), which accounts for the brief, periodic loss
of orbital energy near periastron at a rate governed by the orbital
frequency.  \Ohmic\ dissipation translates to eccentricity damping
through Equation~(\ref{eq:dadt}), with $a$ decaying at a rate of
$da/dt \sim -\sqrt{a(t)}$. The solution to this ODE yields a quadratic
dependence of $a$ on $t$, so that the time for the orbit to formally
shrink from semimajor axis $a$ to zero goes as $\sqrt{a}$. While the
orbital evolution halts earlier, as $a$ approaches $q$ (see
Fig.~\ref{fig:wdb}), when $a\gg q$ we may approximate the time for the
orbit to circularize as
\begin{eqnarray}
\label{eq:tdamp}
\tau_\text{damp} & \ \approx \ &
2\pi 
\rhop \, \left({G\Mstar}\right)^{1/2}\, 
\frac{\muvac\, a^{1/2} q^6}{\Rstar^6\, \Bstar^2}
\efffac^{-1}
\\
\label{eq:tdampex}
\ & \ \approx \ & 160
\left[\frac{\rhop}{\text{\mtsz 5~g/cm$^3$}}\right]\!
\left[\frac{\Mstar}{\text{\mtsz 0.5~$\Msolar$}}\right]^{\!1/2}\!
\left[\frac{\Rstar}{\text{\mtsz 0.5~$\Rsolar$}}\right]^{\!-6}\!
\left[\frac{\Bstar}{\text{\mtsz 5~kG}}\right]^{\!-2}\!
\left[\frac{q}{\text{\mtsz 1 \Rsolar}}\right]^{\!6}\!
\left[\frac{a}{\text{\mtsz 5 \Rsolar}}\right]^{\!1/2}\!
\efffac^{-1}
\ \text{Myr};
\end{eqnarray}
in the lower equation, the power-law indices apply to an asteroid with
an efficiency factor near the maximum, while the numerical reference
values are typical of a magnetic M dwarf, as in
Equation~(\ref{eq:tinspiralexample}) and Table~\ref{tab:typical}.

Our estimates here are intended to provide a simple scaling relation
for \Ohmic\ power dissipation and orbital evolution time scales (e.g.,
Eq.~(\ref{eq:tdampex})). Our result is limited to eccentricities above
0.4 in the case of orbits in the equatorial plane, and tend to
significantly underestimate the damping time as eccentricity falls
below that value. On the other hand, if $i>0$, the damping time is
faster since the asteroid encounters stronger magnetic field
variations when its motion is not limited to the plane perpendicular
to the stellar dipole. Eventually, these orbits circularize and
inspiral. Secular perturbation theory offers a way to give a more
complete description of these more general scenarios
\citep[e.g.,][]{milani1988}. Alternatively, we can use numerical
simulations (e.g., Figure~\ref{fig:wdb}).  Both approaches require an
instantaneous force, as in Equation~(\ref{eq:F}) for low Reynolds
number.

Unlike the circular orbit case, the efficiency factor does not change
significantly as eccentricity damps at fixed $q \ll a$. Another difference
is that circular orbits must be inclined relative to the stellar
dipole moment for \Ohmic\ heating to occur, whereas eccentricity
damping will occur at any orientation relative to the stellar dipole.

The damping time depends strongly on the periastron distance
($\tau_\text{damp}\sim q^6$).  Small asteroids ($\rp < 1$~km) with
high tensile strength may be able to survive close to the host star,
even when $q$ is much less than a Solar radius
\citep[e.g.,][]{kb2017b}.  Closer to the host star, the magnetic field
strength and the efficiency factor for smaller metallic objects are 
potentially much larger.  Orbits for these objects can circularize very 
quickly compared to the 160~Myr time scale in Equation~(\ref{eq:tdampex}).

\subsection{The requisite stellar magnetic field}

Our estimates of orbital decay and damping rates
(Eqs.~(\ref{eq:tinspiral}) and (\ref{eq:tdamp})) are appropriate to
magnetic red dwarfs. Other stars also provide environments where
damping and inspiral may occur.  To explore the possibilities, we
adopt an optimal asteroid size (Eq.~(\ref{eq:roptorbit})) for orbits
near the Roche limit or stellar surface for each of several kinds of
stars, listed in Table~\ref{tab:typical}. We then derive the
strength of the minimum magnetic field required for an orbital decay
time scale of 1~Myr, comparing this value with observations.

\begin{table}[ht]\footnotesize
\caption{\label{tab:typical} 
Typical and/or adopted parameters for various stellar hosts}
\begin{center}
\begin{tabular}{lccccl}
\hline
\hline
\ stellar host & $\Mstar$ & $\Rstar$ & $\Bstar$ & $q$ & comment/references
\\
\hline
red dwarf & $0.5~\Msolar$ & $0.5~\Rsolar$ & 5 kG & 1~\Rsolar & 
\citep{morin2012} \\
Ap star & $2~\Msolar$ & $1.7~\Rsolar$ & 30 kG & 5~\Rsolar & 
\citep[e.g.,][]{elkin2010} \\
white dwarf &$ 0.7~\Msolar$ & $1.4\ \Rearth$ & $10^8$~G & 1~\Rsolar
& \citep{angel1981}
\\
neutron star & $1.4~\Msolar$ & 10~km & $10^{15}$~G & 1~\Rsolar
& \citep{kaspi2017}
\\
\hline
\end{tabular}
\end{center}
\end{table}

\begin{figure}
\centerline{\includegraphics[width=5.0in]{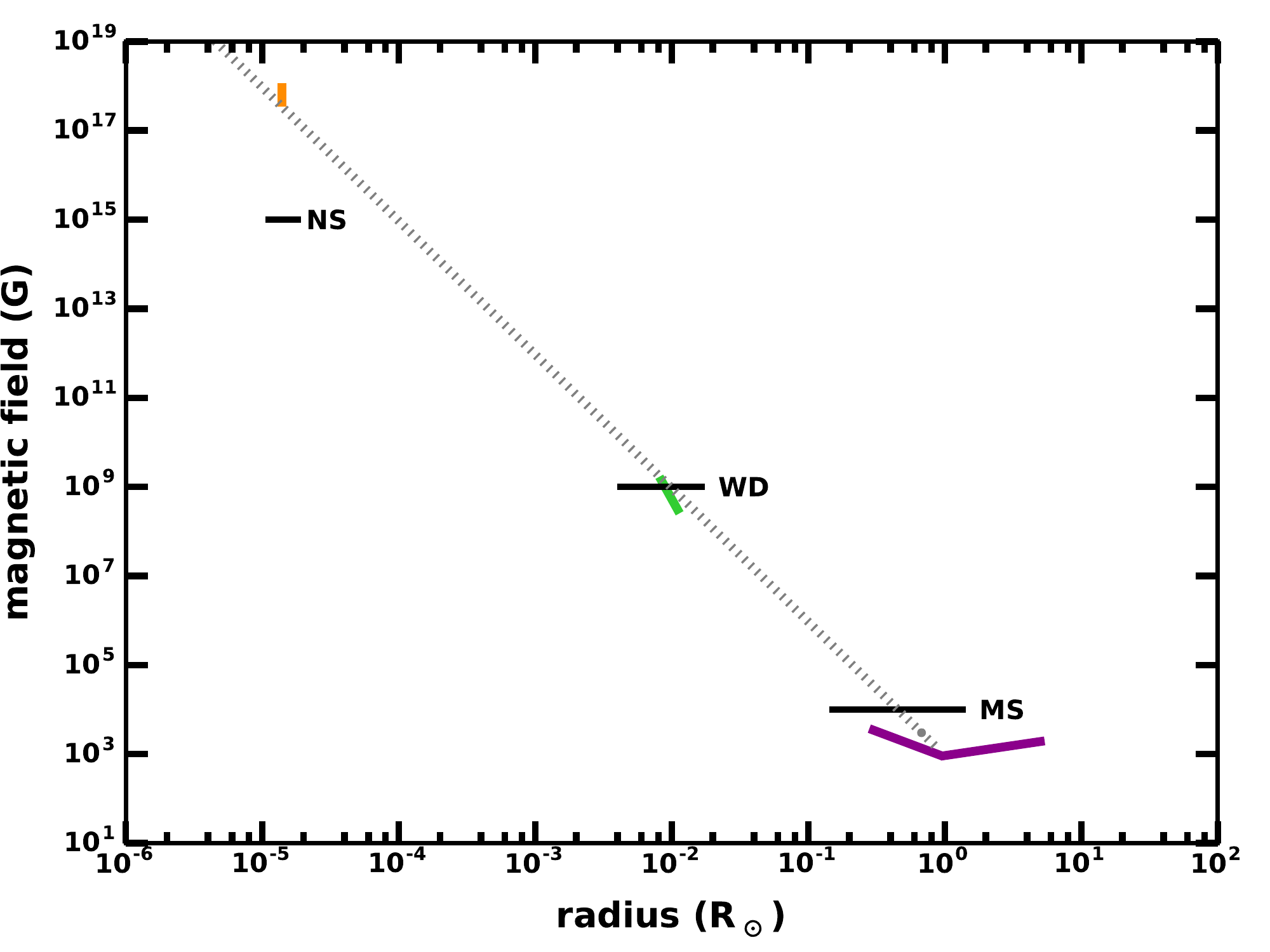}}
\caption{\label{fig:magradius} Comparison between observed magnetic
  field strengths and the field strength that is required to cause
  inspiral of a conducting asteroid within 1~Myr, with focus on.
  neutron stars, white dwarfs, and main-sequence stars.  The
  horizontal lines are the maximum observed magnetic field strengths
  for each of the three types of stars.  The colored line segments are
  the required field strengths for inspiral. For host radii below
  1~\Rsolar, we assume that the asteroid's orbit is at our fiducial
  Roche limit; for larger stars, the orbital distance is at
  1.1~$\Rstar$. Because the inspiral rate has some dependence on
  stellar mass, the required field strengths reflect the mass-radius
  relation. For neutron stars, we assume masses in the range
  1.4--2.5~$\Msolar$, and a fixed stellar radius (10~km).  The dotted
  reference line below a solar radius shows the required field
  strength when the mass of the stellar host is fixed at
  1~$\Msolar$. }
\end{figure}

Figure~\ref{fig:magradius} shows the results. The strongest
observed magnetic fields around late-type main-sequence stars
\citep[$\sim 5\times 10^3$~G;][]{morin2012} and white dwarfs
\citep[$10^9$~G;][]{garcia-berro2016} are high enough to cause orbital
inspiral on megayear time scales.  The field strengths of neutron
stars \citep[$10^{15}$~G;][]{kaspi2017} are insufficient. While the
magnetic field is strong at the surface of a neutron star, it is
comparatively weak near the Roche limit.

To summarize, orbital evolution on a time scale of 1~Myr is possible
around some solar-mass stars if the local magnetic field strength near
the Roche limit is greater than $\sim 50$~G.  For main-sequence stars,
this condition means field strengths at the stellar surface of a few
kilogauss or more. For white dwarfs, surface fields of at least
$10^8$~G are required.
Neutron stars, specifically magnetars, seem promising, but the high
field strengths at the stellar surface are mundane at orbital
distances considered here. We do not give up on neutrons stars,
however, as we discuss in \S\ref{sec:discuss}
\citep[see][]{cordes2008}.

\section{Thermal evolution}%
\label{sec:heat}

Electromagnetically induced eddy currents in a conducting asteroid
produce heat, entirely at the expense of the asteroid's orbital
energy. How the thermal energy from Ohmic heating is distributed
throughout the asteroid depends on its structure and composition.
Furthermore the full thermal evolution of the asteroid must take into
account stellar radiation and radiative losses, as well as the heat
flux within it. We consider all of these effects in this section.

It turns out that thermal evolution time scales for the cases
considered here tend to be fast compared to the orbital time scales
described in \S\ref{sec:orbit}. Thus, while orbital elements change
during thermal evolution (indeed, the latter are driven by the
former), we assume that the changes are small and that \Ohmic\ heating
rates are constant. Relaxing this assumption generally leads to
slightly more rapid thermal evolution than we described here.

To get a sense of the impact of \Ohmic\ heating, we first consider a
simple homogeneous body with heat capacity $\cpc$ and a radius $\rp$
that is optimal for \Ohmic\ dissipation
(Eq.~(\ref{eq:roptorbit}). Given some input power from electromagnetic
induction (Eq.~(\ref{eq:P}) and neglecting other sources and sinks of
energy, the asteroid experiences an average temperature increase at a
rate of
\begin{eqnarray}
\label{eq:DT}
\frac{\Delta T}{\Delta t} & = &   
\frac{3 \Pavg}{4\pi \cpc\rhop\rp^3}\\
\ & \approx & 10 
\left[\frac{\rhop}{\text{\mtsz 5~g/cm$^3$}}\right]^{-1}\!
\left[\frac{\cpc}{\text{\mtsz 0.85~J/K$\cdot$g}}\right]^{-1}\!
\left[\frac{\Mstar}{\text{\mtsz 0.3~$\Msolar$}}\right]^{1/2}\!
\left[\frac{\Rstar}{\text{\mtsz 0.5~$\Rsolar$}}\right]^{6}\!
\left[\frac{\Bstar}{\text{\mtsz 5~kG}}\right]^{2}\!
\left[\frac{q}{\text{\mtsz 1~\Rsolar}}\right]^{-15/2} \efffac 
\ \ \text{K/yr},
\end{eqnarray}
where the numerical values for bulk properties are typical of
terrestrial material \citep{elkins-tanton2011}, while the
astrophysical parameters are based on a close-in circular polar orbit
around a magnetic red dwarf (Table~\ref{tab:typical}).  An asteroid on
an eccentric orbit with a similar closest approach and apoastron at
$\sim$10~\Rsolar\ (0.05~AU) heats up more slowly, at a rate of over
ten degrees a century.

As the asteroid heats up, it will also radiate. When a cold asteroid of 
homogeneous composition has a circular polar orbit, it eventually reaches
a surface temperature of
\begin{eqnarray}\label{eq:Ts}
T_s 
& = &
\left(\frac{\left<\Pohm + \Prad\right>}{4\pi\sigma_\text{SB}\rp^2}\right)^{1/4}
\sim
\left\{\begin{array}{ll}
\rp^{1/4}\Mstar^{1/8}\Rstar^{3/2}\Bstar^{1/2}{a^{3/4}}\,\efffac^{1/4}
\ \ \ \ & (\Pavg\gg \left<\Prad\right>) \\
\Tstar \, (\Rstar/2a)^{1/2} & (\Pavg \ll \left<\Prad\right>),
\end{array}
\right. 
\end{eqnarray}
where the limiting cases in the rightmost expressions apply when the
power from \Ohmic\ heating dominates over the power from stellar
heating ($\Prad$) and visa versa.  Here we assume that the $\Prad$ is
the total stellar flux incident on the asteroid's $\pi\rp^2$ cross
section (its albedo is negligible) and that the asteroid is a black
body radiator.  If we neglect stellar radiation, the temperature of an
optimal-size asteroid around a red dwarf is
\begin{equation}\label{eq:Tsexample}
T_s \approx \left\{
\begin{array}{rl}
   400 \ \text{K} \ \ \ \ & (\text{iron/nickel, $\rp = 3$~km}) \\
   730 \ \text{K} \ \ \ \ & (\text{water, $\rp = 64$~km}) \\
   1600 \ \text{K} \ \ \ \ & (\text{rock, $\rp = 1400$~km}),
\end{array}
\right.
\end{equation}
based on electrical conductivities indicated in
Equations~(\ref{eq:roptmetal})--(\ref{eq:roptrock}).  Solid objects
smaller than the optimal radius radiate interior heat efficiently and
are cooler. Although it is harder for larger objects to radiatively
cool, the \Ohmic\ heating is also less efficient. With $\efffac\sim
1/\rp$ for $\rp > \ropt$, larger bodies are also cooler than asteroids
of optimal size. In the absence of other heat sources, the listed
temperatures are as hot as asteroids get when their conducting cores
are exposed.

The values in Equation~(\ref{eq:Tsexample}) are also representative of
orbits around white dwarfs and Ap stars with parameters listed in
Table~\ref{tab:typical}. The highest temperature listed, 1400~K, is
comparable to an asteroid's equilibrium temperature from heating by 
stellar radiation from the dwarf stars. At least with regard to 
energy flow at the asteroid's conducting surface, \Ohmic\ dissipation is only
barely competitive with radiative heating, and only for low-luminosity
hosts.

\subsection{Heat flow within an asteroid}

In general, \Ohmic\ heating preferentially occurs near the surface of
the conducting core. In the low Reynolds number regime (I), the
angle-averaged current density in a spherical conductor is linear in
distance from the center. With the power density scaling as the square
of the current density, the total \Ohmic\ power loss in thin concentric
shells with radii $r_s$ is proportional to $r_s^4$. Although eddy
currents exist throughout the conductor in this regime, two thirds of
the power is deposited in the outer 20\% of the conducting core. At
higher Reynolds number where the skin effect is important, all the
heating occurs in a thin surface layer; in an iron/nickel core, this
layer has a formal depth of less than a centimeter.

As the surface of the conducting core heats up, the thermal energy
spreads to neighboring regions, according to the thermal diffusivity
of the medium, $\alpha$. For an iron/nickel core, $\alpha =
0.5$~cm$^2$/s \citep[see][]{elkins-tanton2011}; for rocky material,
the diffusivity is lower, $0.03$~cm$^2$/s
\citep[e.g.,][]{gibert2003, sramek2012}.  For any homogeneous,
spherical asteroid, the time scale for spreading heat is
\begin{eqnarray}
\tau_\text{diff} & = & \frac{\rp^2}{\pi^2\alpha}, 
\\
& \approx & 580
\left[\frac{\rp}{\text{\mtsz 3 km}}\right]^2
\left[\frac{\alpha}{\text{\mtsz 0.5 cm$^2$/s}}\right]^{-1}
\ \text{yr.}
\end{eqnarray}
In the lower equation, numerical values are appropriate for an 
iron/nickel sphere.  Kilometer-size metal asteroids reach thermal 
equilibrium quickly compared to orbital evolution time scales. 
Large rocky bodies, with radii of $10^3$~km or more, have diffusion 
times exceeding 1~Gyr. Heat transport in large bodies is faster if 
there are bulk flows or convection, as in the Earth's mantle and 
water worlds \citep[e.g.,][]{munk1966}.

\subsection{\Ohmic\ heating with a non-conducting mantle}

The thermal history of differentiated solids is more interesting.
Asteroids with an iron-rich conducting core and a non-conducting
silicate mantle are common
\citep[e.g.,][]{elkins-tanton2011, neumann2012, scheinberg2015}.
With this structure, the mantle traps heat generated in the core, 
significantly increasing the core temperature from \Ohmic\ heating.

To assess this effect, we assume that an asteroid is made of a
conducting core with radius $\rc$, surrounded by a mantle/crust of
thickness $\Delta r = (\rp-\rc)$, where $\rp$ is the asteroid's full
radius.  Table~\ref{tab:props} shows the bulk properties of the core
and the mantle. In this model, \Ohmic\ heating takes place only in the
core.  While the instantaneous eddy currents from electromagnetic
induction within the core have cylindrical symmetry
(Eq.~(\ref{eq:J})), we assume that on average, heating is evenly
distributed within concentric spherical shells.

\begin{table}[ht]\footnotesize
\caption{\label{tab:props} 
Bulk properties of asteroids
}
\begin{center}
\begin{tabular}{lccl}
\hline
\hline
\ \ \ \ \ name & symbol & fiducial value & comment/references \\
\hline
\multicolumn{4}{l}{\ \ \ \ \ \textit{metallic core:}}
\\
density & $\rho$ & 8~g/cm$^{3}$ & (iron/nickel in differentiated bodies)
\\
electrical conductivity & $\sigma$ & $1\times 10^6$ S/m & \citet{dekoker2012} 
\\
relative permeability & $\murel$ & $5\times 10^3$ & (if $T>\Tcurie \equiv 10^3$~K, 1 otherwise)
\\
thermal diffusivity & $\alpha$ & 0.5--1.0 cm$^2$/s & \citet{elkins-tanton2011}
\\ 
specific heat & $\cpc$ & 0.85 J/g\,K & \citet{bartels1991}
\\
solidus/liquidus & -- & 1213/1233 K & \citet{formisano2013} 
\\
heat of fusion & -- & 250 J/g & (appropriate to iron)
 \\ 
\multicolumn{4}{l}{\ \ \ \ \ \textit{rocky mantle/crust:}}
\\
density & $\rhom$ & 3~g/cm$^{3}$ & 
\\
electrical conductivity & $\sigma_m$ & $0.01$ S/m &  \citet{xu1998}
\\
thermal diffusivity & $\alpham$ & 0.03-1.0 cm$^2$/s & 
\citet{opeil2010}
\\
specific heat & $\cpm$ & 0.80 J/g\,K & \citet{ghosh1998}
\\ 
solidus/liquidus & -- & 1425/1850 K & \citep{formisano2013} 
\\
heat of fusion & -- & 400 J/g & \citet[][silicates]{ghosh1998}
\\ 
\hline
\multicolumn{4}
{p{.7\textwidth}}{\scriptsize
If a parameter has no specified symbol, then it is used in our
code but not in the text.
A parameter that is not listed (e.g., conductivity of rocky crust material) 
has a null value.}
\end{tabular}
\end{center}
\end{table}

When the heat flow through the asteroid reaches a steady state, the
energy flux through the mantle equals the \Ohmic\ power dissipated in
the core.  In a simple model where the thermal conductivity of the
mantle is constant, the core temperature is
\begin{eqnarray}
\label{eq:Tc}
T_c & \sim & \frac{\Pavg\fmantle}{4\pi\rhom\cpm\alpham\rc}
+ T_s
\end{eqnarray}
$\rhom$, $\cpm$, and $\alpham$ are the density, specific heat, and
thermal diffusivity of the mantle, respectively, and $\fmantle =
\Delta r/\rp$ is the mantle's thickness relative to the full
radius. The surface temperature of the asteroid, $T_s$ (in
Eq.~(\ref{eq:Ts})), depends on the stellar radiation, but also on the
rate of the heat transport between the core and the surface.  On a
close-in orbit, the high heating rate and low thermal diffusivity of
rock conspire to give formal values of $T_c$ that can exceed tens of
thousands of degrees even with negligible power input from starlight.

As an example, an asteroid with a 3~km iron/nickel core on a close-in
circular polar orbit around a red dwarf (Table~\ref{tab:typical})
reaches a temperature around 400~K (Eq.~(\ref{eq:Tsexample})) if the
core is exposed. An insulating mantle of only 0.15~km thickness formally
allows the core temperature to reach more than 25,000~K. However, as the core 
is a ferromagnet, it will not reach this temperature. Instead, as the 
core temperature rises through the Curie point, around 1000~K, the 
permeability and the efficiency factor for \Ohmic\ heating drop. The core 
then settles to a steady state with $T_c$ near the Curie point, 
depending on the details of how $\murel$ depends on temperature and 
the influence of stellar radiation on such a close-in orbit. 

If the asteroid had a substantially thicker insulating blanket, then
the prediction for the core temperature would lie significantly above
the Curie point. With a mantle that is 3~km thick, the same as the
core radius, the formal core temperature at steady state would be over
9000~K. As this example shows, trapping of heat by a crust or mantle
allows for high core temperatures in conducting asteroids with a wide
range of sizes and composition, even if they are not of optimal size
for \Ohmic\ heating.

In a more realistic scenario, the mantle that blankets an asteroid's
\Ohmic-heated core is dynamic, with regions nearer the core that melt
and transport heat efficiently by convection instead of molecular
diffusion. Efficient heat flow within the mantle cools the core,
transporting energy toward the asteroid's surface where it can be
radiated away along with the heat from stellar radiation. We include
these effects in the examples we provide next.

\subsection{Thermal evolution of a large asteroid: two examples}

To show how \Ohmic\ heating plays out in an astrophysical
context, we choose a large, differentiated asteroid like 4~Vesta, on a
close-in, eccentric orbit around a 0.5~\Msolar~red dwarf with an
effective temperature of 3500~K (Table~\ref{tab:typical}). The
asteroid's orbit has a periastron distance of 1~\Rsolar\ and semimajor
axis of 5~\Rsolar, about 0.005~AU and 0.025~AU, respectively. The
asteroid's physical radius is $\rp = 250$~km, and its metallic core radius
$\rc$ is half that size.  Table~\ref{tab:props} lists bulk
properties. We assume that the asteroid acquires its orbit in a cold
state with a ferromagnetic core, and that it heats up through the
Curie temperature and melting points, maintaining a high electrical
conductivity in the core and relatively low conductivity elsewhere.

We estimate the thermal evolution from the 1-D spherical thermal
diffusion equation, solving it with a finite difference code, as in
\citet{hevey2006}. At the outer boundary, the surface temperature of
the asteroid is set to balance input of thermal energy flow from the
interior of the asteroid and from starlight of the host with loss from
blackbody radiation. At the inner boundary, the temperature gradient
is zero.  Energy input from eddy currents depends on the magnetic
Reynolds number. For a ferromagnetic iron/nickel core, $\Rm$ is high
(regime III); energy is deposited in a single finite-difference bin.
If $\Rm$ is low (regime I), the power is distributed as across bins as
$r^4$.  We set time steps using the Courant condition, and also
iteratively smooth the sharp discontinuity in the thermal conductivity
between the core and the mantle with a box car function as needed for
numerical stability.

The code tracks the phase of matter and accounts for heat of fusion as
solids go through melting, similar to the algorithm of
\citet{elkins-tanton2011}. The transition across the Curie point is
assumed to be sharp, with $\murel$ modeled as a step function.
When the temperature in a bin exceeds the melting point, we increase
the thermal diffusivity linearly with temperature between solidus and
liquidus from the thermodynamic value to $\alpha = 1.0$~cm$^2$/s as a
way to mimic convective heat flow \citep{hevey2006, sahijpal2007},
although this choice may underestimate the effect
\citep[.e.g][]{neumann2014}. We assume that there is no mixing between
a liquid core and a liquid mantle.

\begin{figure}
\centerline{\includegraphics[width=5.0in]{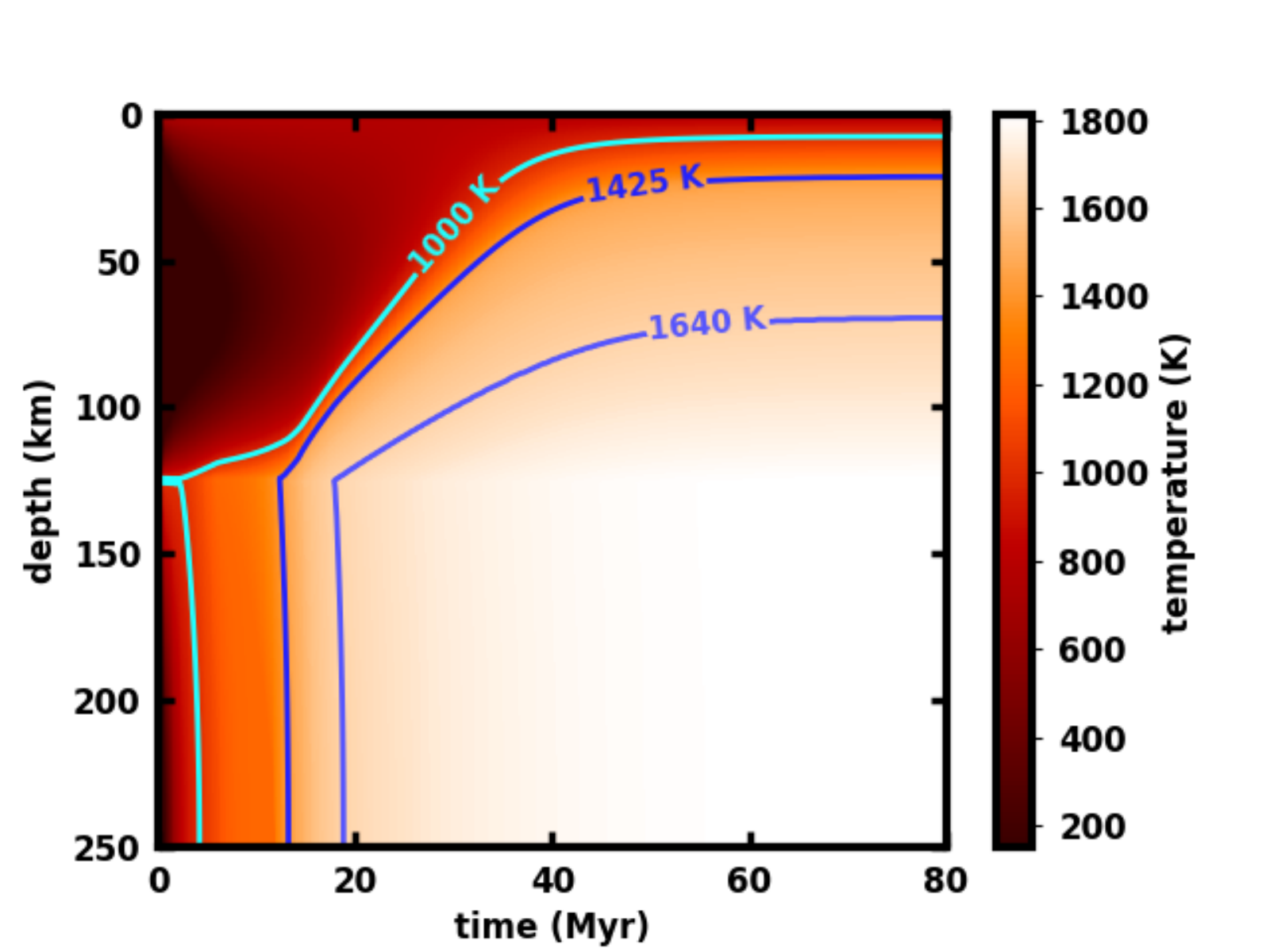}}
\caption{\label{fig:meltxmetalconvectecc} The thermal evolution of a
  250~km differentiated asteroid with an iron/nickel core and a rocky
  mantle on an eccentric orbit around a red dwarf. This density plot
  shows the temperature as a function of depth from the surface
  (vertical axis), and time (horizontal axis). The ``isotherm'' at
  1425~K corresponds to solidus for the mantle material; the contour
  at 1625~K is the 50\%\ melt temperature. The cyan contour indicates
  our adopted Curie temperature, applicable to an Fe/Ni core.  Stellar
  radiation also contributes to the overall thermal evolution. Its
  impact is evident from the increase in temperature near the
  asteroid's surface over the first $\sim 10$~Myr. }
\end{figure}

Figure~\ref{fig:meltxmetalconvectecc} illustrates the thermal
evolution of the Vesta-like asteroid on an eccentric orbit around its
red dwarf host. The plot shows temperature as a function of depth from
the asteroid's surface and how it changes with time.  We assume that
the asteroid is initially cold; as indicated in the Figure, the core
temperature rapidly heats up beyond our adopted Curie point of
1000~K. The mantle starts melting after about 15~Myr, and forms a
region of partial melt that extends to a crust of about 20~km
thick. Stellar radiation also contributes, helping to maintain a
surface temperature above 800~K, close to the equilibrium temperature
for the asteroid's eccentric orbit.
  
\begin{figure}
\centerline{\includegraphics[width=5.0in]{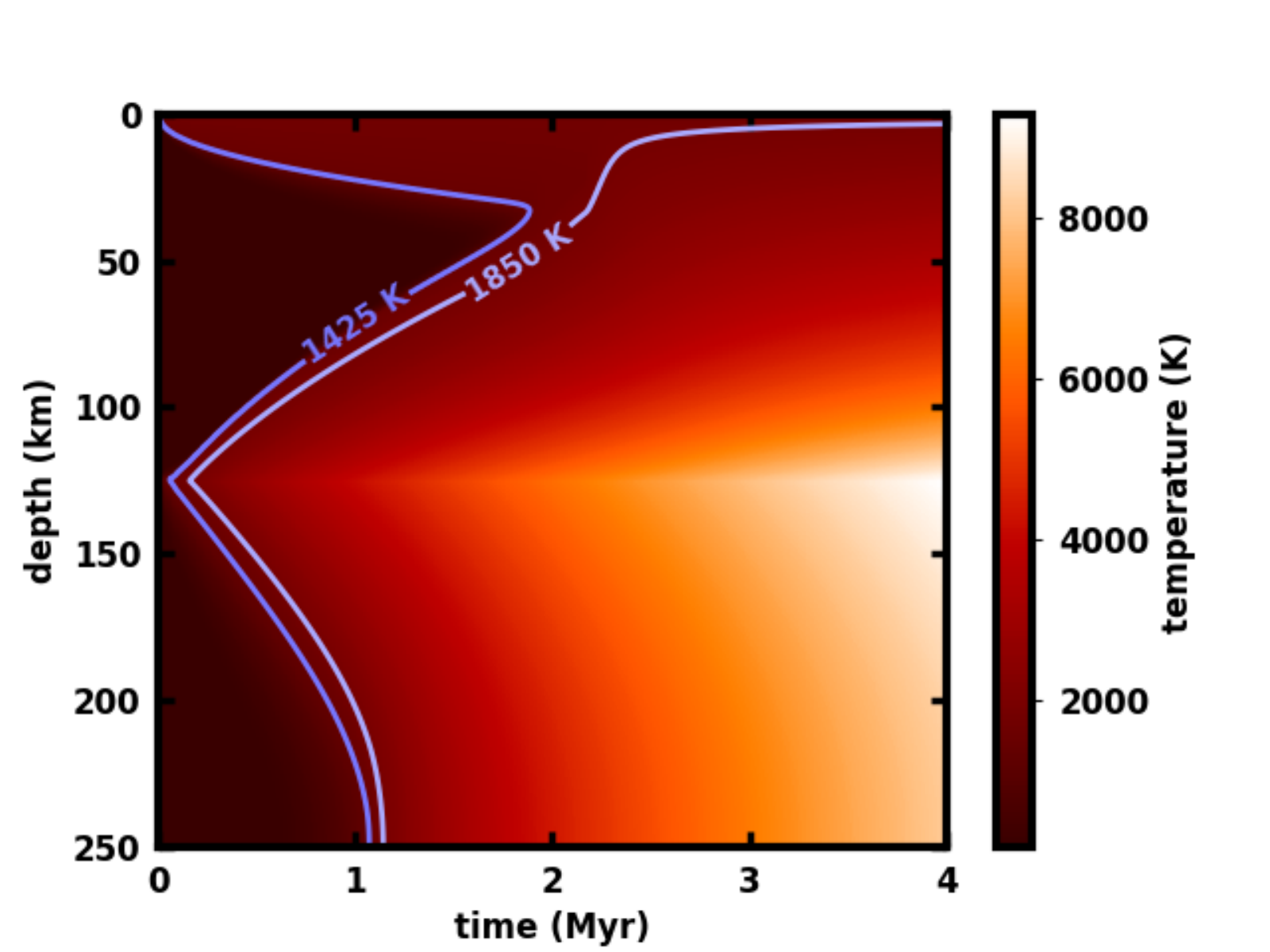}}
\caption{\label{fig:meltxmetalconvectcirc} The thermal evolution of a
  250~km differentiated asteroid with an iron/nickel core and a rocky
  mantle on a close-in circular orbit (0.005~AU) around a red
  dwarf. This density plot shows the temperature as a function of
  depth and time, as in Figure~\ref{fig:meltxmetalconvectecc}. Here,
  we show isothermal contours for the solidus (1425~K) and liquidus of
  the mantle (1850~K). In this case, the \Ohmic\ heating rate is high
  and the core temperature rises considerably above liquidus. Because
  the asteroid is close its stellar host, its equilibrium temperature
  with is close to liquidus. Thus, the asteroid is nearly completely
  melted except for a thin semi-solid crust.}
  \end{figure}

We explore variations on the scenario depicted in
  Figure~\ref{fig:meltxmetalconvectecc}.  If the Vesta-like asteroid
  were to achieve a circular orbit at that same perihelion, the core
  could experience runaway \Ohmic\ heating.
  Figure~\ref{fig:meltxmetalconvectcirc} provides an illustration
  where $T_c$ grows well above 10,000~K. While the state of matter at
  these temperatures is unclear to us, the configuration does not seem
  stable.  The role of stellar radiation in the thermal evolution is to
  quickly bring the surface temperature to just below liquidus. The
  asteroid thus becomes a lava sphere, with an extraordinary amount of
  thermal energy building up in its core.

The role of stellar radiation is notable in both of the above
examples.  For the asteroid on an eccentric orbit, for which the
equilibrium temperature is cool (about 780~K), stellar heating causes
the convective melt zone to extend radially outward, thereby
increasing the efficiency of heat transport from the core. Although
starlight is a significant part of the energy budget, it has only
modest impact on the core temperature. For the asteroid on a close-in
orbit, starlight alone can heat the surface rock almost to liquidus.
Furthermore, because the mantle is almost fully melted --- a
homogeneous medium in our model --- the equilibrium temperature adds
to the steady-state core temperature (Eq.~(\ref{eq:Tc})). However,
\Ohmic\ heating is so strong that the contribution from stellar
radiation is not important.

When the core temperatures are above solidus, volcanism is a possible 
outcome \citep[e.g.,][]{wilson1991, taylor1993}. A modest amount of
volatile elements in the mantle could drive explosive eruptions
\citep{wilson1991}, providing obscuring clouds of dust and a mechanism
for metals from the asteroid to pollute the stellar surface. In an
evolved asteroid, it is unclear how much volatile material might have
survived earlier epochs of heating or whether the hard crust is too 
thick for liquid to break through to the surface
\citep[e.g.,][]{elkins-tanton2011}. If crust thickness is the main
issue, the two scenarios depicted in
Figures~\ref{fig:meltxmetalconvectecc} (cool, thick solid crust) and
\ref{fig:meltxmetalconvectcirc} (hot, essentially a complete melt)
likely bracket conditions for volcanism.

To summarize this section, the thermal history of asteroids
experiencing \Ohmic\ heating is rich and includes the formation of a
deep magma ocean and runaway heating of the core. We caution that the
idealization of a two component (conducting$+$non-conducting) asteroid
may be naive. For example, in the 250~km differentiated asteroid in
Figures~\ref{fig:meltxmetalconvectecc} and
\ref{fig:meltxmetalconvectcirc}, we ignore that the mantle becomes a
magma ocean with conductivity that might be significant.  If $\sigma =
1$~S/m or more \citep[e.g.,][]{gaillard2005} this ocean can
inductively heat even more efficiently than the metal core.  Other
interesting physics may be missed by our use of a 1-D thermal
diffusion solver, as when an asteroid is tidally locked to the stellar
host. Temperatures would build up to higher levels and in more
localized regions than in our calculations.

\section{Discussion}\label{sec:discuss}

Guided by the estimates provided in the previous sections, we examine
how electromagnetic induction and \Ohmic\ heating might play out for
several types of conducting bodies (\S\ref{subsec:bodies}) around
different kinds of stellar hosts (\S\ref{subsec:hosts}).

\subsection{The orbiting bodies}\label{subsec:bodies}

\mysubsubsection{Cold iron/nickel cores}

On an orbit with periastron of 1~\Rsolar\ around a solar-mass star,
the optimal size of a cold, ferromagnetic iron-nickel core is about
3~km.  This core is capable of eccentricity damping on time scales of
hundreds of millions of years around the stellar hosts considered here
(Table~\ref{tab:typical}). The damping time could be significantly
smaller, if those hosts have stronger magnetic fields or if the
periastron distance is inside the nominal Roche radius
(Eq.~(\ref{eq:tdamp})).  Once settled close to the host star, stellar
radiation alone causes the asteroid's core temperature to rise above
the Curie point.  Then, \Ohmic\ heating no longer operates efficiently
(the core is in regime III, $\Rm \gg \murel^2 \sim 1$). Heat input and
orbital evolution slow.

If a heated core manages to melt the surrounding mantle, it is
possible that eddy currents form in the magma ocean instead. If the
magma conductivity were $\sim$1~S/m, then this scenario would lead to
efficient \Ohmic\ heating and orbit evolution if the surface of the
ocean were at a radius of roughly 150~km. A smaller size would be
optimal if the asteroid were on an orbit were closer than a Solar
radius to its host star.

\mysubsubsection{Hot metallic asteroids or meteors}

A cold, small iron/nickel asteroid, $\rp \lesssim 1$~km will orbitally
damp slowly, likely not on interesting time scales.  However, if the
asteroid is on a close-in orbit and its temperature rises above the
Curie point, then the efficiency of \Ohmic\ heating increases. The
optimal radius of an orbiting metallic paramagnetic sphere is roughly
150~m. In this case, heating is possible, but cooling is efficient,
too. The meteor will inspiral through the Roche limit to be tidally
shredded and accreted by the star.

\mysubsubsection{Water worlds}  

At an optimal size of $\sim 70$~km, a water world is capable of
circularizing to a close-in orbit around a magnetic host star. Once on
a tight orbit, however, the temperature of a water world may easily
rise above the boiling point from \Ohmic\ heating or stellar
radiation. The water then evaporates.  Simple estimates suggest that a
water world receives enough energy from the light of a red dwarf or
white dwarf to vaporize a kilometer-deep layer per year. A
water world will not last long on a close-in orbit.

\mysubsubsection{Rocky asteroids and dwarf planets}  

With electrical conductivity in the range of $10^{-4}$~S/m to 1~S/m,
rocky bodies have an optimal radius between $\sim 100$~km (high
$\sigma$) and $\sim 10,000$~km (low $\sigma$). An object with the
intermediate value of 1400~km ($\sigma = 0.01$~S/m) could qualify as
a dwarf planet. Despite its size, it can experience orbital damping on
a time scale of $\sim 100$~Myr, and inspiral within $\sim 1$~Myr. If the
planet had an insulating crust with low thermal diffusivity compared
to the bulk, then the interior could heat beyond the melting point on Myr-time
scales, comparable to orbital evolution times. In Vesta-size
bodies with radii below optimal for \Ohmic\ heating, a melted interior
could raise the efficiency $\efffac$, since the optimal radius of a magma
ocean is around 150~km.

\subsection{The stellar hosts}
\label{subsec:hosts}

\mysubsubsection{Red dwarfs}

Because of their combination of strong surface magnetic field, stellar
radius and mass, magnetic red dwarfs provide a venue for \Ohmic\ heating
of asteroids by electromagnetic induction.  Numerical values for time
scales for orbital damping, inspiral, and heating given above are
tuned for typical magnetized K and M stars; These processes can be
amplified or accelerated by moving the periastron distance $q$ to well
within the nominal Roche limit (Eq.~(\ref{eq:rroche})), or by choosing
a host with a stronger surface magnetic field strength and/or larger
radius. For example, moving $q$ from 1~\Rsolar\ to 0.8~\Rsolar\ shortens the
damping time by a factor of four and the inspiral time by a factor of
five. Increasing the magnetic field strength from 5~kG by 50\%
\citep[near the upper limit of observed values][]{johns-krull2009}
decreases these time scales by another factor of two.

\mysubsubsection{Other main-sequence or pre-main-sequence stars}

T~Tauri stars and magnetic Ap stars have comparable stellar radii
($\sim 2$~\Rsolar) and strong surface magnetic field strengths (up to
1--30~kG). Inspiral and eccentricity damping times for these objects
are short compared to red dwarfs because of their greater magnetic
field strength and the faster orbital speeds at comparable distance.

Around HD~215441 \citep[Babcock's (Ap) star;][]{babcock1960} with a
field strength of 34~kG, an optimal-size asteroid or dwarf planet with 
periastron $q = 5$~\Rsolar\ (about 0.02~AU) and apoastron of five times 
that distance has a damping time of about 190~Myr.  Moving the
periastron distance inward to 3~\Rsolar\ reduces the damping time to
under 10~Myr. 

T~Tauri stars have weaker fields and make their transition to the main
sequence quickly compared with the eccentricity damping time scale. 
Nonetheless, inspiral on a close-in orbit is also possible around
these stars.

\mysubsubsection{White dwarfs}

Of the $\sim$10\%\ of white dwarfs that are magnetic, roughly half
have field strengths above our fiducial value of $10^7$~Gauss
\citep[Fig.~8 therein]{ferrario2015}. About a fifth of these high
field-strength stars --- about 1\% of all white dwarfs --- have
$\Bstar = 10^8$~Gauss or higher. At this field strength, we expect
similar damping, inspiral and heating time scales as for red dwarfs.
For the rare objects with $\sim 10^9$~G fields, the time scales all
drop by a factor of 100. With the fiducial stellar parameters in
Table~\ref{tab:typical}, based on WD~1145$+$017 \citep{veras2017}, the
eccentricity damping time of an optimal-size asteroid is under 10~Myr
if periastron is near the nominal Roche limit and apoastron is within
about 0.1~AU. 

Contributing to electromagnetic induction around a white dwarf is the
magnetic flux variation when the magnetic dipole moment is not aligned
with the star's spin axis. Magnetic white dwarfs have a rotation
period that ranges from under an hour to over several days, with an
average of just under one day \citep{kawka2007, brinkworth2013,
  kawaler2015}.  For comparison, the orbital period at
1~\Rsolar\ around a 0.7~\Msolar\ star is about 3.3 hours.  While there
is a tendency for the stars with the stronger fields to have rotation
periods that are much longer than average \citep{ferrario2005},
stellar rotation may impact \Ohmic\ heating of bodies around it.

How ``spin-orbit coupling'' plays out depends on the alignment between
the spin and the dipole, the rotation rate, and the configuration of
the asteroid's orbit. In some scenarios (slow rotation or perfect
dipole-spin alignment) stellar rotation has little impact. In other
scenarios, \Ohmic\ heating is suppressed when the asteroid's orbit
corotates with a misaligned magnetic dipole. When the stellar
rotation is fast compared to the orbital angular speed, or the orbit
counterrotates relative to a spinning dipole, \Ohmic\ heating is
magnified. We consider this possibility separately for neutron stars.

\mysubsubsection{Neutron stars}

With radius of about 10~km, a mass of 1.4~\Msolar, and a magnetic
field strength of $10^{15}$~G at the stellar surface, a neutron star
has extraordinary potential for \Ohmic\ heating. However, the
magnetic field strength is less than 5~G near the Roche limit,
which is an order of magnitude weaker than around a dwarf star at
comparable distance. Electromagnetic induction can still be important
for close-in asteroids, but the origin of magnetic flux variation is
stellar rotation, not orbital motion \citep{cordes2008,
  kotera2016}. Thus we focus only on the thermal evolution of the
orbiting body.

Neutron stars have a range of spin rates and magnetic field strengths.
Millisecond pulsars have frequencies of $f = \omega/2\pi \sim
10$--1000~s$^{-1}$ \citep[the pulsar PSR~J1748-2446ad has the fastest
  spin rate, with $f = 716$~Hz;][]{hessels2006} and magnetic field
strengths around $10^8$~G \citep[e.g.,][]{mukherjee2015}.  Magnetars
have dipole fields as strong as $10^{15}$~G and spin rates of seconds
\citep[see][and references therein]{kaspi2017}.  Since efficient
\Ohmic\ heating scales as $B^2\omega$
(Eq.~(\ref{eq:PI}))\footnote{Unlike \citet{kotera2016}, we ignore
  electromagnetic scattering, treating $\epsilon\omega \ll
  \sigma$. For the conductivities discussed here, and for observed
  neutron star spin rates, this assumption remains valid.}, we focus
on magnetars, preferring strong field strengths to high spin rates.
Although the magnetic dipole and angular momentum are roughly aligned
in some magnetars, thereby diminishing the amplitude of field
oscillations \citep[e.g.,][]{vogel2014, an2015}, magnetars still are
the most promising class of neutron stars for electromagnetic
induction and \Ohmic\ heating.

For orbits at 1~\Rsolar, near the nominal Roche limit around a
magnetar, and for a stellar rotation frequency of $f=1$~Hz,
we apply Eq.~(\ref{eq:ropt})) to obtain the
optimal radius of an induction-heated asteroid.
When the asteroid is composed of non-magnetic material, we find that
\begin{equation}
\ropt \sim 1.2 \left[\frac{\sigma}{\text{\mtsz 1~S/m}}\right]^{-1/2}\,
\left[\frac{f}{\text{\mtsz 1~Hz}}\right]^{-1/2}
\ \text{km} \ \ \ \ (\murel \approx 1).
\end{equation}
From this expression, rocky bodies with $\sigma = 0.01$~S/m have an
optimal radius of 12~km, while asteroids with an iron/nickel core
heated above the Curie point, $\ropt$ is about 1.2~m.  In a separate
calculation, we estimate that a cold, ferromagnetic asteroid has an
optimal radius of just over 25~m.

The average heating rate in the oscillating magnetic field is
\begin{equation}
\frac{dT}{dt} \sim 0.9 
\left[\frac{\rhop}{\text{\mtsz 5g/cm$^3$}}\right]^{-1} 
\left[\frac{\cpc}{\text{\mtsz 0.85 J/K$\cdot$g}}\right]^{-1} 
\left[\frac{\Bstar}{\text{\mtsz 10$^{15}$}}\right]^2 
\left[\frac{q}{\text{\mtsz 1~\Rsolar}}\right]^{-6}\! \efffac \ \text{K/yr}.
\end{equation}
At this rate, a 12~km rocky body will reach a steady state around
$\sim 200$~K in the absence of other energy sources. Larger bodies,
with less efficient radiative cooling, reach higher temperatures. A
crust with low thermal diffusivity allows for runaway heating and
melting in the interior of bodies of all sizes.

\subsection{Observational outcomes}

Given the range of asteroid size, composition and stellar
host for which \Ohmic\ heating may be important, we consider 
a set of possible observational consequences of the phenomenon.
We begin with the most promising one.

\mysubsubsection{Pollution of white dwarf atmospheres}

Sinking times for metals at the surface of a white dwarf is so rapid
that no metals should persist in the stellar atmosphere
\citep{dupuis1992, althaus2000, koester2009}.  Yet O, Mg, Al, Si, Ca,
and Fe are abundant in about a third of all DA and DB stars
\citep[e.g.,][]{zuckerman1998, zuckerman2010, gansicke2012,
  koester2014, kepler2015, kepler2016, farihi2016}. There must be a
reservoir of external material available to these stars, presumably of
planetary origin, delivering mass at an average rate of $10^5$~g/s to
$10^{10}$~g/s \citep[e.g.,][]{farihi2016}.

\Ohmic\ heating can result in the delivery of metals to the
atmosphere of a white dwarf. A small metallic or metal-rich asteroid,
scattered onto an orbit within a Solar radius of the stellar host,
can damp and inspiral to be tidally shredded within its Roche
radius. At a periastron distance of ten times the stellar radius, or
about 0.1~\Rsolar, a hot, 10~m metallic asteroid would be near both
its Roche limit and the optimal radius $\ropt$. The damping time then
is thousands of years for orbits with apoastron at 10~\Rsolar\ around a
host with a surface magnetic field of strength of $10^8$~G. Inspiral
is even faster. After tidal shredding, the next steps include
evaporation and gas accretion \citep[e.g,][]{jura2003, jura2008,
  debes2012, veras2014, kb2017c}.

Similarly, kilometer-size rocky bodies would be near both the optimal
radii and their Roche limit at about a half of a Solar radius
\citep[Fig.~1 therein]{kb2017b}. Damping time scales from 10~\Rsolar\ exceed 
1~Myr, but inspiral times are only thousands of years around a
star with $\Bstar = 10^8$~G. Our prediction is thus that electromagnetic
induction can accelerate the flux of planetary material orbiting close 
to magnetic stars as compared with their non-magnetic counterparts.

\mysubsubsection{Dust from explosive volcanism}

Large asteroids or dwarf planets, if near optimal size for Ohmic
heating, can experience interior melting
(Figures~\ref{fig:meltxmetalconvectcirc} and
\ref{fig:meltxmetalconvectecc}) opening up the possibility of
volcanism.  Volcanic eruptions can produce copious amounts of gas and
dusty solids; eruptions on Jupiter's moon Io, driven by heating from
tidal stresses \citep{lainey2009}, contain as much as $10^9$~g in
submicron-size dust \citep{kruger2003, geissler2008}. Terrestrial
volcanos can eject thousands to millions of cubic kilometers of
material, or over $10^{18}$~g if in the form of solids.

The numerous outstanding issues for volcanism include the amount and
composition of ejected material, the number of eruption sites, the
frequency of eruptions, and the ejecta speed relative to the
asteroid's escape velocity. If parameters are favorable for the
release of volcanic dust onto close-in orbits, a consequence could be
short-lived occultation of the stellar host. By ejecting $10^{14}$~g
of micron-size particles an eruption can temporarily dim the
stellar host \citep[see][Eq.~9 therein]{kb2005}. 

A second consequence of volcanic dust production is the delivery of
metals to the stellar host.  With $\sim 10^9$~g of material per event,
daily eruptions are needed to account for the observed metal accretion 
rates in white dwarfs. An annual explosive supervolcano could provide 
a similar metal pollution rate. 

\mysubsubsection{Exploding asteroids?}

\Ohmic\ heating of an asteroid with a modest-size conducting core
($\rc \sim 1$~km) and an insulating crust can lead to melting of its
interior. If this transition happens near the Roche limit, the loss of
material strength allows tidal forces to overwhelm the asteroid,
rapidly liberating $10^{15}$~g of debris. We speculate that the
observational signature would be similar to a catastrophic collision
between two asteroids \citep{kb2005}. The distinguishing feature of a
tidally shredded asteroid is that the dynamical time corresponding to
the debris cloud is fast --- hours, not days or months --- because of
the close-in orbital distance.

As a result of the high temperature build-up in large asteroids ($\rp
\gtrsim 10$~km, as in Fig.~\ref{fig:meltxmetalconvectcirc}), there is
the possibility that very hot core material might mix with a cooler,
volatile-rich mantle or thick crust.  Even a small molar fraction of
volatiles (e.g., 1\%), if heated rapidly to O($10^3$)~K, could cause a
disruptive expansion of gas within the asteroid.  The resulting
explosive event could be as spectacular as it is speculative.

\subsection{Where to look: Candidates for \Ohmic\ heating}

\begin{itemize}

\item High-field magnetic white dwarfs, analogous to WD~1145+017.  The
  handful of white dwarfs with field strengths above $10^8$~G offers
  the most promising environments for \Ohmic\ heating by
  electromagnetic induction \citep[e.g.,][Table 9
    therein]{briggs2015}. Observables would be atmospheric pollution
  by planetary material and, possibly, occultation by dust. The
  non-magnetic star WD~1145+017 shows compelling evidence for both
  phenomena, likely resulting from the tidal disruption of an asteroid
  near its Roche limit \citep[see][and references
    therein]{farihi2018}. It illustrates the promise of observing
  close-in asteroids as well as the challenge of disentangling the
  effects of \Ohmic\ heating from tidal break-up.

\item Magnetic main-sequence stars. \Ohmic\ heating is possible around
  both magnetic red dwarfs and early-type stars.  An induction-heated
  asteroid or dwarf planet that has volcanic activity could produce
  clouds of obscuring dust.  This scenario might be consistent with
  the deep enigmatic occultations of F star KIC 8462852
  \citep{boyajian2016}. It could also explain the apparent
  disintegration of small, ultra-fast period planets around red
  dwarfs, including the ``super-Mercury'' Kepler-1520~b \citep[aka KIC
    12557548;][]{rappaport2012} and K2-22b \citep{sanchis-ojeda2015},
  which are both orbiting at a distance of about 3~\Rsolar\ around
  their hosts.

\item Pre-main-sequence stars. The environment around a T Tauri star
  can be messy, with a vigorous stellar wind, accretion flows, and
  strong magnetic fields. Induction heating of an asteroid that relies
  on current through the circumstellar plasma
  \citep[e.g.,][]{sonett1970} could be supplemented by Ohmic heating
  driven by changing magnetic flux if the orbit passes close to the
  stellar surface.  There is now evidence of planetary material on
  close-in orbits about some young stars, as reported by
  \citet{stauffer2018}. Phased light curves of these sources show
  deep ``scallop shell'' undulations with periods of hours, suggestive of
  close-in circumstellar gas and dust.  The light-curve patterns can
  persist for months, with occasional abrupt changes after flare
  outbursts. While the connection between these features and
  Ohmic heating of an asteroid is not clear, they
  suggest that planetary material can get close to the stellar
  surface.

\item Magnetars. A large asteroid that is induction-heated to the
  point where it can sustain volcanic activity could result in
  episodic bursts of radiation from the host. Volcanic ejecta 
  transferred from the asteroid to the neutron star surface will
  become relativistic when it impacts the surface. We do not speculate
  further except to note that the rest energy in single volcanic
  eruption of $10^9$~g is over $10^{29}$~erg.
\end{itemize}

\section{Conclusion}
\label{sec:conclude}

Here, we describe how the orbits of conducting bodies with radii in
the range of $\sim 1$~m to over 1000~km (``asteroids'') are modified
by electromagnetic induction as they plow through the magnetic field
of a stellar host. We focus on objects of rocky or metallic
composition, based on simple assumptions: all material is homogeneous
and isotropic, with linear response to electromagnetic fields. Our
choices of density, electrical conductivity, permeability, thermal
diffusivity and specific heat are representative of small bodies in
the solar system. We use analytical solutions for the current density
that depend on the magnetic Reynolds number and permeability to track
\Ohmic\ heating, and a 1-D finite difference code to estimate heat
flow within an asteroid.

Our analysis demonstrates that orbital damping and inspiral from Ohmic
dissipation is possible if the perihelion distance is near the Roche
limit ($\sim 1$~\Rsolar) around magnetic stars with typical field
strengths (Table~\ref{tab:typical}). The time scales for these
processes are around 1~Myr to 1~Gyr. Damping and inspiral times can be
much shorter if periastron distances that are well inside the nominal
Roche limit can be sustained, or if the magnetic dipole field strength
were increased. For example, the damping time of an orbit with
apoastron at 0.1~AU around a white dwarf falls to about 10,000 years
if the magnetic field strength is at the upper end of the observed
range ($10^9$~G), and periastron is at 0.3~\Rsolar. The inspiral time
from 0.3~\Rsolar\ is a few decades.

\Ohmic\ heating occurs on even faster time scales. When placed on an
orbit around a magnetic star that generates electromagnetically
induced eddy currents, an asteroid or dwarf planet plausibly heats at
rates as high as a few Kelvin per year. Exposed conducting cores of
$10^3$~km or less reach a steady state at an uninteresting temperature
(e.g., below the melting point, or less than the equilibrium
temperature from radiative heating).  A crust or mantle that traps heat
causes core temperature to formally rise well above the melting
point. 

Our analysis is limited in several ways. First, we consider only the
magnetic dipole field of the host star.  Stellar magnetic fields may
be more complicated \citep[e.g.,][for white dwarfs]{putney1995,
  fendt2000, wickramasinghe2001} and may alter the flux variations
experienced by asteroids on close-in orbits. The magnetic field may
also have a component that arises in the circumstellar environment.
The magneto-rotational instability within a protoplanetary accretion
disk may support field strengths of 100~G or more at distances of a
few solar radii \citep[e.g.,][]{mohanty2018}.  Second, we do not
consider the effect of the electric field induced by the orbital
motion of an asteroid through the (approximately uniform) local
magnetic field, which may be important in the presence of a plasma or
stellar wind \citep[see also \citealt{mottez2011} for a study
  that also includes the back-reaction of the plasma]{sonett1970}. 
  Finally, we neglect the general effect of local field gradients,
  including the force on the induced magnetic dipole of a conducting
  asteroid by the gradient in the stellar field
  \citep[e.g.,][]{ray2018}. Our focus here is just on phenomena
  directly tied to the loss of orbital energy.

Despite these limitations, the results presented here show
possibilities for detecting \Ohmic\ heating of asteroids and dwarf
planets. The effect is most likely to occur in the most extreme cases
of close-in orbits and high magnetic field strengths.  Promising
environments are Ap stars, magnetic red dwarfs, and magnetic white
dwarfs. Despite very strong surface fields, orbit-powered
\Ohmic\ heating is not effective around neutron stars.  Instead, the
spinning magnetic dipole moment of a magnetar is substantially more
effective as a generator of induced currents and \Ohmic\ heat
\citep{cordes2008, kotera2016}.

The primary connections of \Ohmic\ heating to observations include
orbital inspiral of a conducting asteroid through the Roche limit,
delivering a large reservoir of metals to the immediate vicinity of
the stellar host. We are encouraged by recent discoveries of
ultra-short period planets as a sign that such close-in orbits are
achievable \citep{sanchis-ojeda2014}. If an asteroid or dwarf planet
has an insulating crust or mantle, then \Ohmic\ heating can raise the
temperature of core material high enough to potentially cause volcanic
eruptions or large-scale explosions even before tidal break-up. Such
events could deliver the elements observed in the atmospheres of
metallic white dwarfs, and might generate enough dust to dim
main-sequence stars. Around a neutron star, a large volcanic eruption
could send ejecta into the deep, relativistic gravitational well,
beaming news of the event to viewers at vast distances.

\acknowledgements

We are grateful to M. Geller, C. Boehme, O. Starykh, C. Williams, and
Z. Zheng for comments and helpful discussions. We thank an anonymous
referee for providing guidance that significantly improved our
manuscript.  We also thank J. Nagel for key advice on the problem of
induced eddy currents.  NASA provided essential support for this
program through a generous allotment of computer time on the NCCS
'discover' cluster and {\it Outer Planets Program} grant NNX11AM37G.

\bibliography{planets}{}
\bibliographystyle{apj}

\end{document}